\documentclass{aa}  
\usepackage[varg]{txfonts}
\usepackage[dvipsnames]{xcolor}
\usepackage{hyperref}
\let\linenumbers\relax 

\begin{document} 

    \title{Polarimetric insights into a potential binary supermassive black hole system in Mrk~231}

    \titlerunning {A polarimetric investigation of Mrk~231}
    
   \author{J. Biedermann \inst{1,2}\thanks{\href{mailto:julie.biedermann@astro.unistra.fr}{julie.biedermann@astro.unistra.fr}}    
            \and
            F. Marin\inst{1}
            \and
            T. Barnouin\inst{1}}

   \institute{Universit\'e de Strasbourg, CNRS, Observatoire astronomique de Strasbourg, UMR 7550, F-67000 Strasbourg, France
              \and
              Univ. Lyon, Univ. Claude Bernard Lyon 1, CNRS/IN2P3, IP2I Lyon, F-69622, Villeurbanne, France
              }

   \date{Received January 7, 2025; accepted Month Day, 2025}
   
   \abstract
   {Markarian 231 (Mrk~231) is one of the brightest ultraluminous infrared galaxies (ULIRGs) known to date. It displays a unique optical-ultraviolet (optical-UV) spectrum, characterized by a strong and perplexing attenuation in the near-UV, associated with a sudden polarization peak. 
   }
   {The aim of this study is to clarify the puzzling nature of Mrk~231's spectrum by examining the potential existence of a binary supermassive black hole (SMBH) system in its core. To this end, we combined photometric information used in prior studies, with archival polarimetric data to provide a more thorough approach to the quasar's structure and emission mechanisms. In particular, we evaluated the binary SMBH model as a potential explanation for the near-UV cutoff, while exploring its polarization response, for this source, for the first time.
   }
   {Building on previous spectro-photometric modeling, we investigated the hypothesis that the core of Mrk~231 may host a binary SMBH system. In this scenario, the accretion disk of the primary, more massive SMBH is responsible for the optical-UV spectrum. The disk of the secondary, less massive SMBH, would be expected to essentially emit in the far-UV. We applied this model to archival photometric and polarimetric data of Mrk~231 and tried to obtain the best fit possible. To support our findings, we performed radiative transfer calculations to determine the spatial disposition of each main component constituting Mrk~231.
   }
   {We find that a binary SMBH model can reproduce both the observed flux and polarization of Mrk~231 remarkably well. We infer that the core potentially hosts a binary SMBH system, with a primary SMBH of about 1.6 $\times$ 10$^{8} M_{\odot}$ and a secondary of about 1.1 $\times$ 10$^{7} M_{\odot}$, separated by a semimajor axis of $\sim$ 146 AU. The secondary SMBH drives a degree of polarization of $\sim$ 3 \% between 0.1 and 0.2 $\mu$m, with a corresponding polarization position angle of about 134$^{\circ}$, which is consistent with scattering within an accretion disk. The primary SMBH and the structure around it are responsible for a degree of polarization of $\sim$ 23 \% between 0.3 and 0.4 $\mu$m with a corresponding  polarization position angle of about 96$^{\circ}$, that is possibly attributed to scattering within the quasar's wind. Finally, our model predicts the existence of a second peak in polarized flux in the far-ultraviolet, a telltale signature that could definitively prove the presence of a binary SMBH.
   }
   {These results strongly support the hypothesis of a binary SMBH in Mrk~231 and emphasizes the need for new far-ultraviolet spectropolarimeters to clearly detect the existence of subparsec binary SMBHs in nearby quasars.
   }

    \maketitle

\section{Introduction}
\label{introduction}

Markarian 231 (Mrk~231), the nearest quasar with a redshift of $z$ $\sim$ 0.042, is an ultraluminous infrared galaxy (ULIRG) that exhibits a Seyfert~1 optical spectrum \citep{Adams_1972, Sanders_1988}. It is likely represents the final stage in the interaction between two galaxies \citep{Armus_1994}, resulting in strong accretion and hence its quasar's status. Mrk~231 is known for a few observational key signatures, such as the presence of broad absorption lines (BAL) in its optical spectrum ; in particular, Fe~II absorption lines ($\sim$ 8000 km.s$^{-1}$)at 2446 \AA \ and 2676 \AA. Thus, Mrk~231 can be classified within a specific subcategory of BAL quasars, the FeLoBALs, characterized by weak iron ionization \citep{teng_1980,Smith_1995}. Moreover, Mrk~231 exhibits a particularly and remarkable unique ultraviolet (UV)-optical-infrared (IR) spectrum \citep{Smith_1995}. Although its optical-to-IR spectral energy distribution (SED) appears to be typical of quasar and could be explained based on a combination of an active galactic nucleus (AGN) and starburst activity \citep{farrah_2003}, its spectrum shows a pronounced cutoff in the near UV (at wavelengths around 3000 \AA). It also displays an atypical second peak of continuum emission in the far-UV, suggesting a complex physical process taking place in the core of Mrk~231.

This puzzling feature is associated with one of the most notable aspects of Mrk~231, often ignored : an exceptionally high degree of linear polarization around the UV dip, first noted by \citet{Thompson_1980}. The polarization degree reaches notably high values in the near-UV, with $P$ = 18 \% $\pm$ 1.7 \% observed between 2755 and 2797 \AA \ and $P$ = 13.7 \% $\pm$ 0.2 \% between 3350 and 3500 \AA \  \citep{Goodrich_1994,Smith_1995}. From the near-UV to near IR continuum, the polarization of Mrk~231 then exhibits a significant and monotonic decrease, from approximately 20 \% in the UV down to a very low value of 0.5 \% in the IR \citep{Kemp_1977, Smith_2004}. This behavior is extremely unusual, as Seyfert galaxies and radio-quiet quasars typically exhibit an optical linear polarization degree of a fraction of a percent, which remains approximately constant from the UV to the IR band (once the contribution of the host galaxy removed). This polarization peak is also associated to a rotation of the polarization position angle, from $\sim$ 130$^{\circ}$ at $\lambda \approx$ 2200 \AA\, to $\sim$ 80$^{\circ}$ at $\lambda \approx$ 2300 \AA\, and then $\sim$ 100$^{\circ}$ at $\lambda \approx$ 8000 \AA\, \citep{Smith_1995}. 

The strong near-UV attenuation and weak continuum at shorter wavelengths were investigated by \citet{Leighly2016}, who attributed it to circumstellar reddening. In contrast, \citet{Yan_2015} proposed an alternative hypothesis, suggesting the existence of a milliparsec binary supermassive black hole (SMBH) system having a significant impact on the UV part of the spectrum. In their scenario, the low UV emission is dominated by the lower mass SMBH (4.5 $\times$ 10$^6 M_\odot$), which accretes matter via a thin, small disk that principally emits in the UV. On the other hand, the higher mass SMBH (1.5 $\times$ 10$^8 M_\odot$) has a low accretion rate and emits radiation through an advection-dominated accretion flow (ADAF). A circumbinary disk surrounds these two SMBHs, emitting the optical and IR photons. The authors hypothesized that the abrupt optical decline towards the UV is a characteristic of such a binary SMBH system and the observed cutoff in the UV waveband flux is probably due to a gap in the ADAF, opened by the secondary SMBH which migrates inside the circumbinary disk. The abrupt rise towards the farUV would then be due to the inner edge of the smaller accretion disk. A unscaled schematic of the model is shown in Fig.~1 of \citet{Yan_2015}.

This hypothesis is appealing since Mrk~231 is in an advanced stage of galactic merger, as previously written and is therefore a strong candidate to host a binary system of SMBHs \citep{Armus_1994}. Thus, we considered how we could determine whether this hypothesis is correct. One way is to rely on polarimetry, as it has been proven multiple times that it is a powerful tool for investigating the inner geometry of obscured or spatially unresolved systems; in addition, it has been used to rejected several binary SMBH candidates (see, e.g. \citealt{Marin2023}). Mrk~231 has been extensively observed in the past, notably due to the aforementioned puzzling polarization peak, but never in the context of a binary SMBH. This is crucial because polarization (both in degree and angle) provide two more independent observables to the total flux, making the constraints much stronger in the case of a model capable of fitting all these multi-technical observation data.

The main goal of this paper is to check whether the data on the total and polarized fluxes could fit within the binary SMBH frame by developing an analytical binary SMBH model for Mrk~231 and studying its total and polarized fluxes signatures. In Sect.~\ref{Compilation}, we present our methodology for compiling and analyzing all the available polarimetric data, including an exhaustive search within the archives and a new reduction of UV data obtained by the Hubble Space Telescope (HST) Faint Object Camera (FOC). In Sect.~\ref{Modeling}, we describe our construction an analytical model based on the work of \citet{Yan_2015} with respect to a milliparsec binary SMBH system in the core of Mrk~231. This model was used to reproduce the SED of a single accretion disk and then updated to integrate additional components in order to produce a realistic SED model that incorporates polarization predictions. We compared the model to the observation and try to get the best set of parameters to reproduce the data. In Sect.~\ref{STOKES}, we explain how we used radiative transfer modeling using the \textsc{stokes} code in order to simulate the complex interactions in the core of Mrk~231 and determine whether our analytical solution is supported by state-of-the-art simulations. Section~\ref{Discussion} provides a critical analysis of the results, especially in the context of papers that reject the binary SMBH hypothesis. Finally, in Sect.~\ref{Conclusions}, we summarize the main results of our paper. These results strongly support the binary SMBH hypothesis, as it successfully reproduces both the total flux and polarization features observed for Mrk~231 without invoking exotic physics. 

\section{Compiling all polarimetric data}
\label{Compilation}

\subsection{Archival search}
\label{Compilation:archives}

This study is based on existing archival data\footnote{\url{https://ui.adsabs.harvard.edu/}} that are freely available. We specifically searched the website database for flux measurements of Mrk~231 that had corresponding polarimetric observations. We discarded all papers with purely photometric or spectroscopic results, since our goal in this work is to collect simultaneous total and polarized flux measurements. We found 14 papers that matched our basic selection criterion and the final catalog of polarimetric observations of Mrk~231 is presented in Table~\ref{tableaux_archive_mrk231}. The table is structured as follows: column (1) lists the reference of the study, (2) the instrument and telescope used, (3) the spectral band of the observation, (4) the aperture of the telescope in arcseconds, and (5) whether an image of Mrk~231 is present in the archives. 

\begin{figure}
    \centering
    \includegraphics[width=1\linewidth, trim={0.15cm 0.5cm 0cm 0.4cm}, clip]{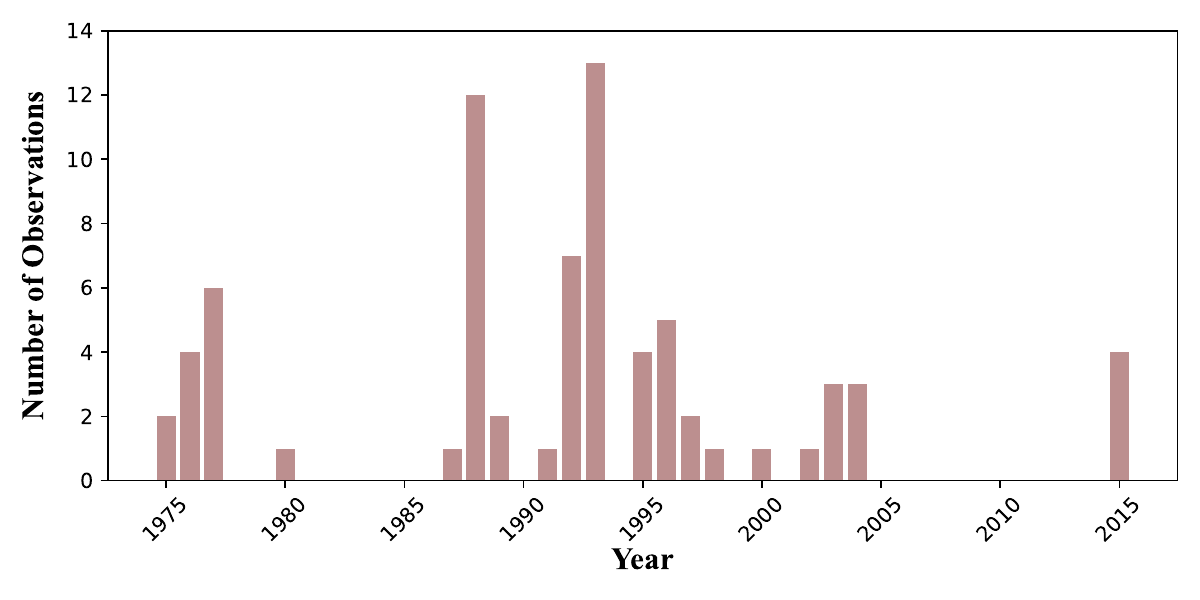}
    \caption{Number of observation per year for Mrk~231.}
    \label{nb_obs_mrk231}
\end{figure}

\begin{figure}
    \centering
    \includegraphics[width=1\linewidth, trim={0cm 14cm 13cm 15cm}, clip]{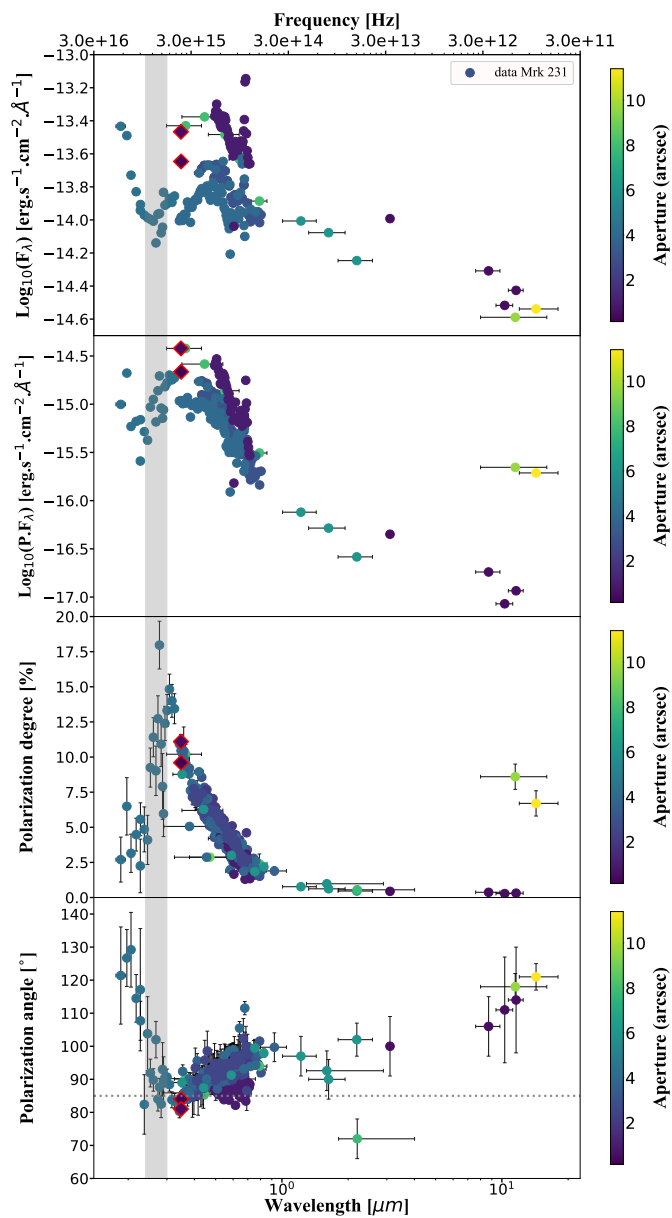}
    \caption{\textit{Top panel} : Reddening-corrected total flux of Mrk~231. \textit{Second panel} : Polarized flux of Mrk~231, established from the multiplication of the flux and the polarization degree. \textit{Third panel} : Linear polarization degree. \textit{Bottom panel} : Polarization position angle of Mrk~231. The aperture is color-coded (in arcseconds). The gray zone represents the waveband in which we observe the cut-off in the UV spectrum, associated with a strong rise in the degree of polarization and a distinctive rotation in the polarization position angle. The data used to construct those plots are the one summarized in Table~\ref{tableaux_archive_mrk231}. The data highlighted in red as the new ones we derived from the poorly explored HST/FOC observation of Mrk~231.}
    \label{polarimetric_data_mrk_231}
\end{figure}

In Fig.~\ref{nb_obs_mrk231}, we show the temporal distribution of the number of polarimetric observations, which are listed in Table~\ref{tableaux_archive_mrk231}. We find that the publications cover about 40 years (1975-2015, based on the observation dates). No new polarimetric observations have been achieved in the last decade. The amount of observational data is not as large as, for example, in the case of NGC~1068 \citep{Marin2018b} or NGC~4151 \citep{Marin2020}; however, these data are sufficient to construct a full UV-to-IR SED. The time coverage of the data also allowed us to look for variability in polarization. From Fig.~\ref{polarimetric_data_mrk_231} , we can clearly see that there is none (within the measurement uncertainties). 

As shown in Table~\ref{tableaux_archive_mrk231}, the measurements were taken using various instruments across different wavebands, with various apertures and weather conditions. Nevertheless, as already shown in \citet{Marin2018b}, such a discrepancy is not prohibitive in such research, as long as the polarization properties of the source are not variable over time (and those of Mrk~231 are not). The data points collected in our catalog are shown in Fig.~\ref{polarimetric_data_mrk_231}. It gives the reddening-corrected flux, the polarized flux (that is the multiplication of the total flux with the polarization degree), the degree of linear continuum polarization, and the angle of polarization as a function of wavelength, with the aperture used for the observation color-coded accordingly. However, before we analyzed these SEDs, we were able to determine a "new" measurement of Mrk~231 polarization at 3480 \AA.

\begin{table*}[ht!]
    \caption{Catalog of published polarimetric measurements of Mrk~231.}
    \centering
    \resizebox{\textwidth}{!}{
    \renewcommand{\arraystretch}{1.5}
    \begin{tabular}{ p{38mm} p{90mm} p{62mm} p{28mm} p{12mm} }
        \hline 
        \large{Reference}
        & \large{Instrument}
        & \large{Waveband}
        & \large{Aperture}
        & \large{Imaging}
        \\
        \hline
        \citet{Gallagher_2005}
        & spectropolarimeter SPOL Bok 2.3-m telescope \newline Faint Object Camera, Hubble Space Telescope
        & 4200-8200 \AA , B, R filter       \newline 3480 \AA
        & 3"                                \newline 0.47"
        & False                             \newline True
        \\
        \citet{Goodrich_1994}
        & dual-beam spectropolarimeter 3 m Shane reflector, Mount Hamilton
        & 4419 - 7190 \AA
        & 14"
        & False
        \\
        \citet{Jones_1989} 
        & Minnesota infrared Polarimeter (MIRP) Two Holer photopolarimter UM/UCSD 1.5-m telescope Mount Lemmon Observatory
        & 1.2 , 1.65, 2.2 $\mu$m \newline  U, B, V filters \newline 0.64 $\mu$m, 0.79 $\mu$m
        & 6" \newline 6" \newline 8"
        & False \newline False \newline False
        \\
        \citet{Kemp_1977}
        & 2.25-m telescope Steward Observatory
        & 1.6 $\mu$m \newline 2.2 $\mu$m
        & 5.9" \newline 7.8"
        & False \newline False
        \\
        \citet{Lopez_2017}
        & MMT-Pol, AO secondary system on the 6.5-m MMT, Arizona \newline Canari-Cam 10.4-m Gran Telescopio CANARIAS
        & 3.1 $\mu$m filter \newline 8.7 $\mu$m, 10.3 $\mu$m, 11.6 $\mu$m
        & 0.5" circular \newline 0.5" circular
        & True \newline True
        \\
        \citet{Martin_1983}
        & Steward Observatory 2.3-m telescope \newline Las Campanas 1.0 and 2.5-m telescope \newline Kitt Peak 1.3 and 2.1-m telescope \newline University of Western Ontario 1.2-m telescope
        & 3800-5600 \AA
        & 9"
        & False
        \\
        \citet{Schmidt_1985}
        & ITS spectropolarimeter 3-m Schane telescope Lick Observatory
        & 3400-7100 \AA
        & 4" circular
        & False
        \\
        \citet{Siebenmorgen_2001}
        & ISOCAM instrument - Infrared Space Observatory
        & 12 $\mu$m \newline 14.3 $\mu$m
        & 9.6" \newline 11.4"
        & False \newline False
        \\
        \citet{Smith_1995}
        & Faint Object Spectrograph Hubble Space Telescope \newline Steward Observatory Kitt Peak 2.3-m telescope
        & 1575 - 3294 \AA \newline 4000 -7800 \AA
        & 4.3" \newline 4"
        & False \newline False
        \\
        \citet{Smith_2004}
        & ISIS dual-beam spectrograph 4.2-m William Herschel Telescope (WHT)
        & 4900 - 7200 \AA 
        & 1"
        & False
        \\
        \citet{Thompson_1980}
        & University of Western Ontario (UWO) filter polarimeter \newline - Kitt Peak National Observatory 1.3-m telescope \newline - Kitt Peak National Observatory 2.1-m telescope \newline - UWO 1.2-m telescope \newline Oke multichannel spectrophotometer (MCSP) \newline - Hale Observatories 5.1-m telescope \newline University of California at San Diego (UCSD) Digicon detector \newline - Steward Observatory 2.3-m telescope
        & \hfill \newline 4040-4800 \ \AA, 5550-6230 \ \AA, 7100-7950 \ \AA \newline 3200-3850 \ \AA, 8000-8600 \ \AA \newline \hfill \newline \newline 3400 - 10\,500 \AA \hfill \newline \newline 3900 - 7000 \AA
        & \hfill \newline 5.2" \newline 6" \newline 7.7" \hfill \newline \newline 3.6" \hfill \newline \newline 2.5"
        & \hfill \newline False \newline False \newline False \hfill \newline \newline False \hfill \newline \newline False
        \\
        \citet{Thompson_1988}
        & Steward Observatory 2.3-m telescope \newline Las Campanas 1.0-m and 2.5-m telescope \newline Kitt Peak 1.3 and 2.1-m telescope \newline University of Western Ontario 1.2-m telescope
        & 2900 - 5500 \AA\,, 2340 - 7580 \AA\,, 5550 - 8480 \AA
        & 4" 
        & False
        \\
        \citet{Ulvestad_1999}
        & Very Long Baseline Array (VLBA) \newline Very Long  Array (VLA)
        & 1.4 to 22.2 GHz \newline 1.5 GHz, 4.8 GHz, 15.0 GHz
        & 4" \newline 1"
        & True \newline True
        \\
        \citet{Young_1996}
        & 3.8-m United Kingdom Infrared Telescope (UKIRT) University of Hertfordshire optical/IR polarimeter HATPOL
        & U, B, V, R, I, J, H, K filters
        & 5"
        & False
        \\
        \hline 
    \end{tabular}
    }
    \tablefoot{The first column is the reference paper, the second column is the instrument and telescope used for the measurement, the third column is the waveband or filters used for the observation, the fourth column is the instrument aperture (in arcseconds) and the fifth column indicates if polarimetric images were taken.}
    \label{tableaux_archive_mrk231}
\end{table*}

\subsection{HST/FOC data reduction}
\label{Compilation:FOC}

During our archival search, we found references to a HST/FOC polarimetric observation of Mrk~231 without any associated images \citep{Gallagher_2005}. This is most uncommon, since the FOC was precisely designed to obtain high resolution polarization images of astronomical sources. Only a single value of the polarization was reported : 15.3\% at 97$^\circ$. Because we have access to a new reduction pipeline for the HST/FOC (see \citealt{Barnouin2023}, \citealt{Barnouin2024} and \citealt{Marin2024}), we decided to investigate the matter and provide new polarimetric data for this source, taking advantage of the final calibration files for the whole FOC dataset published a decade, following the analysis of \citet{{Gallagher_2005},{Kamp2006}}. 

Imaging polarimetry of Mrk~231 was obtained on 1998, November 28, using the FOC on board the HST (observation ID 6444). The detector was used in 128x128 readout format to mitigate saturation. The observation was made through the F346M filter, with a centered wavelength 3480 \AA\, and a bandwidth of 434 \AA\,, along with the three usual polarizer filters. As already mentioned, the first analysis of Mrk~231's FOC polarization was presented in \cite{Gallagher_2005}, where the authors reported a 0.47" aperture polarization degree and angle of $P = 15.3\%$ and $\theta_P = 97^\circ$ (independent on the aperture radius). They argued that the imaging data did not show evidence of extended structure, but without publishing the aforementioned images. Alternatively, \cite{Leighly2016} recovered an observation of the same target through the filter F210M (center wavelength 2180 \AA\, with 164 \AA\, bandwidth) with no polarizer filter, obtained with the same proposal as \citet{Gallagher_2005}, to perform spatial analysis of the HST/FOC image of Mrk~231. They reported a 2D Gaussian fit for the PSF of Mrk~231 with a best-fit value of $0.0491 \pm 0.0016$ arcsec but, again, without showing any image.

\begin{figure}
    \centering
    \includegraphics[width=1.\linewidth,trim={0 0 0 0},clip]{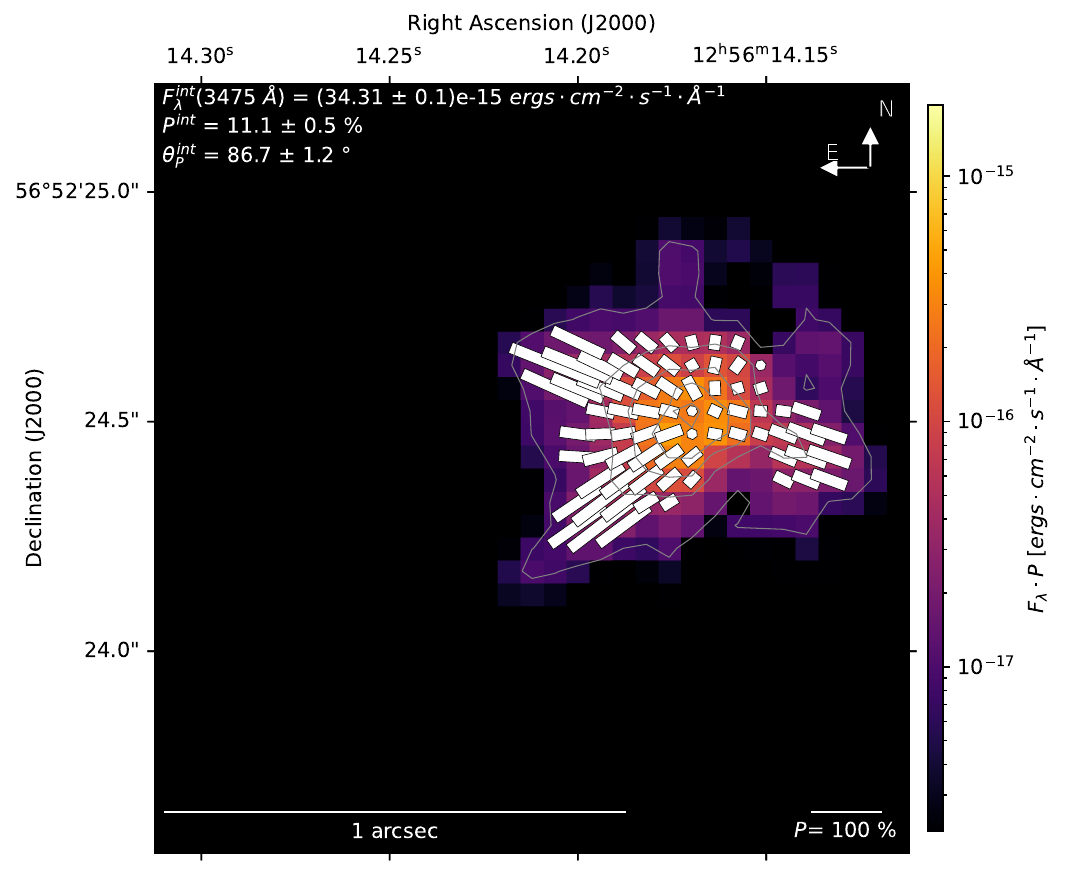}
    \caption{HST/FOC near-UV polarization map of Mrk~231. The electric vector polarization angles are displayed for a signal-to-noise ratio in polarization degree greater than 5 (north corresponding to 0$^\circ$). Values in the top left are integrated over the whole field of view.}
    \label{fig:FOCmap}
\end{figure}

We thus performed a HST/FOC polarization map reduction following \citet{Barnouin2023} and re-sampled the images to the scale of the observed PSF by \cite{Leighly2016} (i.e., 0.05 arcsec). The images were smoothed with the combination algorithm and a gaussian of full width at half maximum of 1.5 times the pixel size, that is, 0.075 arcsec. See all technical details about this reduction is \citet{Barnouin2023}. The obtained polarization map is presented in Fig.~\ref{fig:FOCmap}, overlaid on the polarized flux. We report integrated values on the 1.8''x1.8'' field of view for the total flux density, $F_\lambda(3475$\AA$) = (34.31\pm 0.10) \times 10^{-15}$ ergs cm$^{-2}$ s$^{-1}$ \AA$^{-1}$, debiased polarization degree, $P = 11.1 \pm 0.5 \%$, and polarization angle, $\theta_P = 86.7 \pm 1.2^\circ$. For a circular aperture of radius 0.5'' centered on the peak of polarized flux density, we obtained $F_\lambda = (34.16 \pm 0.09) \times 10^{-15}$ ergs cm$^{-2}$ s$^{-1}$ \AA$^{-1}$, $P = 11.1 \pm 0.5 \%$ and $\theta_P = 86.0 \pm 1.1^\circ$. This region clearly dominates the signal over the whole field of view and these debiased values are in agreement with those reported in \cite{Gallagher_2005}, accounting for more sources of uncertainties. For a circular aperture of radius 0.14'' centered on the peak of polarized flux density, we report $F_\lambda = (22.59 \pm 0.08) \times 10^{-15}$ ergs cm$^{-2}$ s$^{-1}$ \AA$^{-1}$, $P = 9.6 \pm 0.6 \%$, and $\theta_P = 89.0 \pm 1.7^\circ$. These values have been added to the measurements obtained in the archival search and are displayed as red diamonds on Fig.~\ref{polarimetric_data_mrk_231}. We notice a significant variation in flux measurements between 0.3 and 0.8 $\mu$m. As our archival data was obtained with different instruments and apertures, and given the absence of detailed information on the exact observing conditions for each data set, we chose to retain all measurements without applying aperture or normalization corrections.

However, we must note that the full polarimetric observation in native resolution display elongated features that differ from one polarization filter to the other (see Fig.~\ref{fig:FOC_obs}). This is clearly an artifact that is nonphysical, which is probably why the previous authors did not show it. Any depiction of the global polarization pattern in Mrk~231 is thus impossible, since extended structures are affected by these artifacts. We discuss the importance of such artifacts in Appendix~\ref{app:FOCreduc}, but we note that the results we report carefully avoid these artifacts by focusing on the central pixels. Indeed, the obtained polarization map is dominated by the unresolved point source corresponding to the AGN, but no conclusions can be drawn from the observation of the extended emission. 

\subsection{Polarization spectra}
\label{Compilation:spectra}

The wavelength-dependent total flux, polarized flux, linear continuum polarization degree, and angles of Mrk~231 are presented in Fig.~\ref{polarimetric_data_mrk_231}. These spectra range from 0.1 (far-UV) to 20 $\mu$m (mid-IR). Most of the observations were achieved with instrument apertures smaller than 4 arcseconds, with only a few having apertures larger than 8. The largest apertures will often include a significant fraction of the host starlight, while the smallest apertures will better isolate the AGN. 

Focusing on the reddening-corrected total flux first, we can observe the famous break in the SED of Mrk~231, a feature that is extremely uncommon for AGNs. From the far-UV to about 2275~\AA, the flux sharply drop to reach a minimum value (highlighted in gray on the spectrum), then rises again in the near-UV, up to 4580~\AA, and then slowly decreases with increasing wavelengths. The flux of the AGN is slightly variable in time (as all AGNs), explaining the minor discrepancies between flux measurements at the same wavelength and with similar apertures. We note high flux levels at the smallest apertures for a series of measurements that could have caught the source in a bright state. However, apart from the UV break that remains enigmatic, the optical to IR SED of Mrk~231 is rather aligned with the averaged SED of quasars \citep{Saccheo2023}.

The polarized flux of Mrk~231 (second panel) shows a rather different shape than the total flux spectrum. At the wavelengths where there is the UV break in total flux, the polarized flux is strongly rising. From the mid-UV to the near-IR, the overall shape of the polarized flux SED is reminiscent of a multi-temperature black body emission. The only exception to this is in the far-UV, where the polarized flux rises instead of decreasing. There is a break in the polarized SED of Mrk~231, but it is situated at about 2200~\AA, which is about 500~\AA \ less than in the total flux spectrum. 

The shape of the polarized flux is directly related to the behavior of the continuum, linear polarization degree. The observed polarization of Mrk~231 is unique in the zoology of AGNs, with a polarization degree of about 2.5\% in the far-UV, then a sudden and sharp increase of $P$ up to almost 20\% at about 3000~\AA\,, followed by a smooth diminution in the optical band, down to a single percent or less in the IR. Our two HST/FOC points offer a nice fit with the general trend. Despite the large range of different apertures, instruments, and date of observations, the polarization degree spectrum of Mrk~231 is remarkably stable in terms of shape and time. 

Finally, the polarization position angle (last panel of Fig.~\ref{polarimetric_data_mrk_231}) shows an interesting behavior : the electric vector polarization angle varies smoothly as a function of wavelength. It starts at about 130$^\circ$ in the far-UV, then rotates to about 80$^\circ$ in the mid-UV, rises from the near-UV to the optical up to about 100$^\circ$, then decreases again in the near-IR before rising up to 120$^\circ$ in the mid-IR. The polarization angle varies gradually over the wavelength band, without abrupt discontinuities (i.e., each point in the spectrum is consistent within uncertainties with the previous one), indicating that the polarization variations we see from this object are certainly intrinsic. We can compare the polarization angle to the radio jet axis, that was measured to be $\sim$ 5$^{\circ}$ by the Very Long Baseline Array (VLBA, see \citealt{Ulvestad_1999}). We find that the polarization angle is perpendicular to the radio jet axis around 3000 - 5000~\AA, but it deviates from orthogonality before and after. It is a hint that would suggest perpendicular scattering of core photons onto polar winds, as commonly observed for type-2 AGNs \citep{Antonucci1993}.

\section{Reproducing the observation with a toy model}
\label{Modeling}

Our first goal is to reproduce the total SED of Mrk~231 and then analyze the polarimetric responses of this model to determine whether the Mrk~231 data are consistent with a binary SMBH model. This investigation is prompted by the assumptions proposed by \citet{Yan_2015}. 

\subsection{Modeling the SED of an accretion disk}
\label{Modeling:disk}

The model we developed for this study is based on textbook physics, drawing on the work of \citet{Yan_2015}. We intended to simulate the multi-temperature black body emission of an accretion disk and then couple two models to form a (non-interacting) binary system of SMBHs, where each SMBH is surrounding by its own accretion disk. To do so, we considered a Newtonian disk model, which is geometrically thin and optically thick, as described by \citet{shakura_Sunyaev}. Consequently, we have been able to postulate that the disk’s surface emits locally like a black body, a process governed by Planck’s law \citep{Lorenzin}. The black body radiation intensity at a given wavelength is defined in Eq.~\ref{intensity_black_body} :

\begin{equation}
    I_{\lambda}(T) = \frac{2 h c^{2} }{\lambda_{i}^{5}} \frac{1}{ \left(e^{ \frac{h c}{ \lambda_{i} k_{B} T} } -1\right)},
    \label{intensity_black_body}
\end{equation}

with $c$ as the speed of light, $h$ the Planck constant, $k_{B}$ the Boltzmann constant, $T$ the temperature of the emitting ring and, $\lambda_i$ the wavelength for the ring radius, $i$. Within this framework, the temperature at any given point on the accretion disk can be computed as a function of its radial distance from the center. The temperature profile across the disk is derived using the formula presented in Eq.~\ref{accr_disk_temperature} \citep{Pringle_1981} :

\begin{equation} 
    T(r) = \left(\frac{3GM_{*}\dot{M}}{8 \pi \sigma_{s} r^{3}}\right)^{\frac{1}{4}}  \left(\frac{r_{int}}{r_{*}}\right)^{-\frac{3}{4}}\left(1-\sqrt{\frac{r_{*}}{r_{int}}}\right)^{\frac{1}{4}} \label{accr_disk_temperature}.
\end{equation} 

Here, $r_{*}$ is the Schwarzschild radius (the horizon radius of a Schwarzschild black hole, calculated with the formula 2$GM$/c$^2$), $G$ the gravitational constant, $M_{*}$ the mass of the black hole, $\dot{M}$ the mass accretion rate, and $\sigma_{s}$ the Stefan-Boltzmann constant. Each ring of the disk emits a black-body spectrum according to its local temperature as shown by Eq.~\ref{intensity_black_body}.

Now that we have a disk with an inner radius and an emission profile at each radius, we need to determine where does the disk stops. To do so, we used the Toomre parameter : we assumed that the disk remains stable until the accretion rate, $\dot{M}$, defined by Eq.~\ref{critical_accretion_rate}, surpasses the critical accretion rate, $\dot{M}_{+}$, at this radius, as defined in Eq.~\ref{m_croix} \citep{lacosta_2016}. At this specific radius, the disk is no longer stable and ends. These equations are expressed as : 

\begin{equation}
    \dot{M} = \alpha p \dot{M}_{Eddington} \ [g.s^{-1}],
    \label{critical_accretion_rate}
\end{equation}

where $\alpha$ is the viscosity, $p$ is the accretion parameter and $\dot{M}_{Eddington}$ is the Eddington accretion rate, which is defined by Eq.~\ref{accre_rate_edd} :

\begin{equation}
    \dot{M}_{Eddington} = \frac{L_{Eddington} }{\eta c^{2}} = 1.5 \times 10^{18} \frac{M_{*}}{M_{\odot}} \ [g.s^{-1}].
    \label{accre_rate_edd}
\end{equation}

Here, $\eta$ is the radiative efficiency ($\eta = 0.1$, \citealt{lacosta_2016}), $M_{*}$ the mass of the black hole, and $L_{Eddington}$ the Eddington luminosity, with $M_{\odot} = 1.99 \times 10^{33}$ g. The aforementioned critical accretion rate $\dot{M}_{+}$ is defined as : 

\begin{equation}
    \dot{M}_{+}(r) = 8.07 \times 10^{15} \alpha^{-0.01}_{0.1}\left(\frac{r}{10^{10} [cm]}\right)^{2.64}\left(\frac{M_{*}}{M_{\odot}}\right)^{-0.89},
    \label{m_croix}
\end{equation}

where $\alpha_{0.1}$ represents the viscosity parameter normalized to $0.1$, $r$ is the radius of the accretion disk and $M_{*}$ the mass of the black hole. 

\begin{figure*}
    \centering
    \includegraphics[width=1\linewidth, trim={12.5cm 8cm 10cm 15cm}, clip]{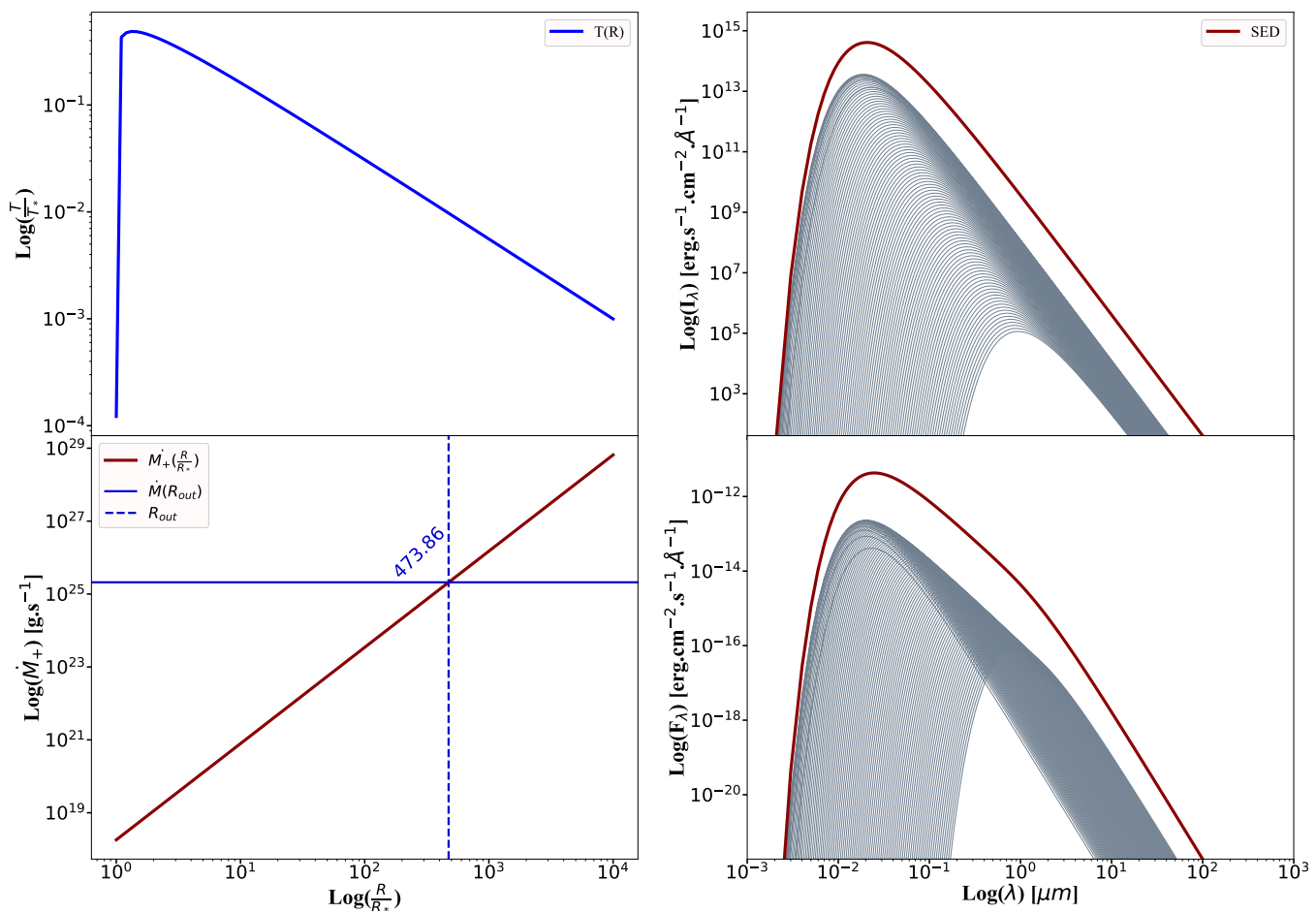}
    \caption{\textit{Top left: } Temperature profile normalized by the black hole temperature $T_{*}$ as function of the radius of the accretion disk (normalized by the Schwarzschild radius of the black hole). \textit{Bottom left: } Critical accretion rate of the accretion disk as a function of the radius of the disk (normalized by the Schwarzschild radius of the black hole). \textit{Top right: } Emitted intensity as a function of the wavelength for each of the 100 disk radius (in gray) and summed up (in red). \textit{Bottom right: } Theoretical flux as a function of the wavelength for a SMBH  with $M_{*,p}$ = 1.5 $\times$ 10$^{8} M_{\odot}$ and an accretion disk with inner radius of $r_{int,p}$ = 3.38 $\times$ 10$^{-3}$ pc and an outer radius of $r_{out,p}$ =  6.80 $\times$ 10$^{-3}$ pc (all 100 radii are shown in gray and summed up in red).}
    \label{sed_simple_system}
\end{figure*}

In this study, we first considered a single SMBH with a mass of  1.5 $\times$ 10$^{8} M_{\odot}$ surrounded by an accretion disk that emits according to the Planck’s law (detailed in Eq.~\ref{intensity_black_body}). Initially, the accretion disk extends from 1 to 10\,000 gravitational radii. We plot the temperature profile of the accretion disk using Eq.~\ref{accr_disk_temperature}, which is also illustrated in the top left panel of Fig.~\ref{sed_simple_system}. The calculation yields a critical accretion rate of $\dot{M}_{+}$(r) = 2.09 $\times$ 10$^{25}$ [g.s$^{-1}$]. We thus identified the intersection between the logarithm of the critical accretion rate as a function of $log(\frac{R}{R_{*}})$ as the value of the accretion rate that surpasses the critical accretion rate. This allowed us to fix the outer radius of the accretion disk, as shown in the bottom left panel of Fig.~\ref{sed_simple_system}. Finally, we employed Eq.~\ref{intensity_black_body} to calculate the SED of the accretion disk of the SMBH as a function of the logarithm of the wavelength. We then constructed our first SED model with the following free parameters : the mass of the SMBH ($M_{\odot}$), the viscosity ($\alpha$), the accretion parameter ($p$), and the system's inclination angle ($\theta$). On the other hand, we imposed a distance between the source and the observer, based on the distance to Mrk~231 equivalent to $z$ = 0.04152 (or approximately 581 millions light years). The simulation of 100 accretion disks radii (in gray lines) allows us to build the summed up SED of the whole disk, represented by the red line in the bottom right panel of Fig.~\ref{sed_simple_system}.

\subsection{Modeling the SED of binary system of SMBH}
\label{Modeling:binarySMBH}

The next step in our study is to develop a SED model for a binary SMBH system by incorporating a secondary SMBH into our existing model. In the modeling of the SMBH binary system at the core of Mrk~231, specific considerations are given to the structures of the accretion disks of the primary and secondary SMBHs. \citet{Yan_2015} proposed adding a second SMBH with a lower mass than the primary SMBH, namely, $M_{*,s}$ = 4.5 $\times$ 10$^{6} M_{\odot}$, so that it orbits the first. In their model, the secondary SMBH is assumed to be close to the primary black hole, within the inner radius of the primary accretion disk.

The outer radius of the accretion disk surrounding the secondary, lighter SMBH is calculated on the basis of the Roche lobe radius, which depends on the mass ratio $q$ of the two SMBHs, as proposed by \citet{Yan_2015}. The Roche lobe radius $R_{RL}(q)$ is given by Eq.~\ref{roche_radius}, which takes into account the semi-major axis $a_{BBH}$ and the mass ratio $q$ to determine the influence of the secondary SMBH \citep{Eggleton_1983} as follows: 

\begin{equation}
    R_{RL}(q) \approx \frac{0.49 \ a_{BBH} \ q^{\frac{2}{3}} }{0.6 \ q^{\frac{2}{3}} + log(1+q^{\frac{1}{3}})}.
    \label{roche_radius}
\end{equation}

The outer radius $r_{out,s}$ is then a fraction $f_{r,s}$ of the radius of the Roche lobe, as shown by Eq.~\ref{R_ext_bh_secondaire} :

\begin{equation}
    R_{out,s} = R_{RL}(q) . f_{r,s}.
    \label{R_ext_bh_secondaire}
\end{equation}

The inner radius of the secondary SMBH is equal to the ISCO radius for a non-rotating black hole (3$\times$ ${2GM_{*,s}}/{c^{2}}$), as it marks the last stable orbit before matter falls into the black hole. On the other hand, the inner radius of the accretion disk around the primary SMBH, $r_{int,p}$, is determined from the Hill radius, $R_{Hill}$, the semimajor axis, $a_{BBH}$, and the mass ratio, $q$, which defines the region in which the primary SMBH exerts a dominant gravitational influence on the accretion disk matter. The inner radius of the primary SMBH, corresponding to the inner edge of the circumbinary disk, is described by Eq.~\ref{R_int_bh_primaire} : 

\begin{equation}
    r_{int,p} = R_{Hill} + \frac{a_{BBH}}{(1+q)},
    \label{R_int_bh_primaire}
\end{equation}

with $a_{BBH}$ as the the semimajor axis. The Hill radius is defined in Eq.~\ref{Hill_radius} as 

\begin{equation}
    R_{Hill} \sim a_{BBH} \left(\frac{M_{*,s}}{3M_{*,p}}\right)^{\frac{1}{3}}.
    \label{Hill_radius}
\end{equation}

Here, $M_{*,s}$, is the mass of the secondary SMBH and $M_{*,p}$ is the mass of the primary SMBH. The outer radius of the primary SMBH is calculated from the Toomre parameter, as explained in the previous section. 

The initial parameters of our model include the mass of the primary and secondary SMBHs ($M_{*,p}$ and $M_{*,s}$ respectively), the inclination angle for each SMBH ($\theta_p$ and $\theta_s$), their viscosity ($\alpha_{p}$ and $\alpha_{s}$), their accretion parameter ($p_{p}$ and $p_{s}$), the semimajor axis ($a_{BBH}$), and the mean Roche radius ($f_{r,s}$). \citet{Yan_2015} did not specify the viscosity ($\alpha$) and accretion parameter ($p$) in their model ; thus we assumed standard values ($\alpha = 0.1$ and $p = 1 $). To plot the SED of each accretion disk for the two SMBHs, we used the same method described in the previous section to find the outer radius of the disk of the primary (more massive) SMBH. 

\begin{figure}
    \centering
    \includegraphics[width=1\linewidth, trim={0.5cm 6cm 5cm 7cm}, clip]{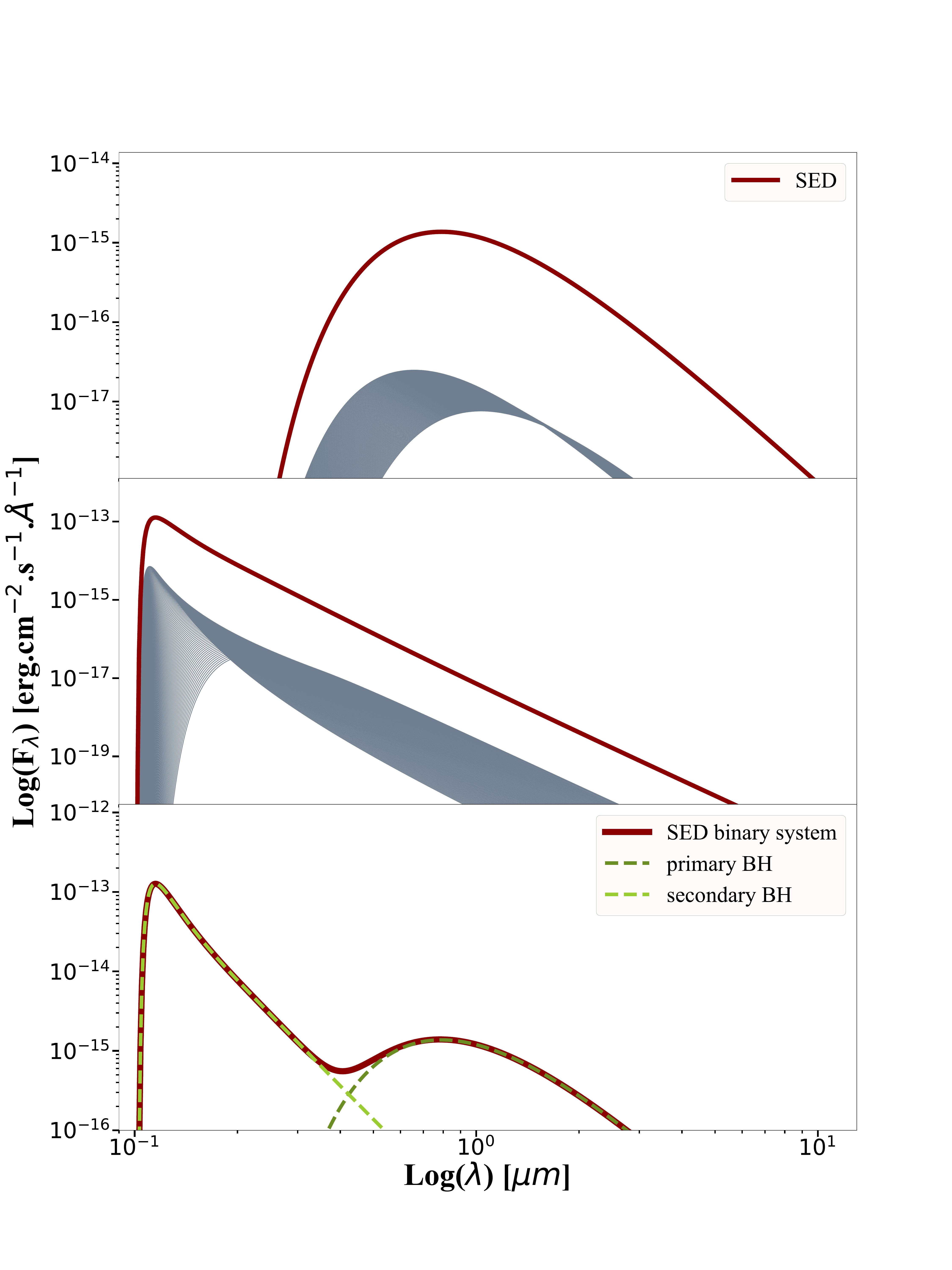}
    \caption{\textit{Top:} SED of the primary (more massive) SMBH in red. Each of the 100 emitting radii are shown in gray. \textit{Middle:} Same plot, but for the secondary (lighter) SMBH. \textit{Bottom:} Combined SED of the two SMBHs in a binary system based on the model proposed by \citet{Yan_2015}, with $M_{*,p}$ = 1.5 $\times$ 10$^{8} M_{\odot}$, $r_{int,p}$ = 3.38 $\times$ 10$^{-3}$ pc and $r_{out,p}$ =  6.80 $\times$ 10$^{-3}$ pc for the primary SMBH, and $M_{*,s}$ =  4.5 $\times$ 10$^{6} M_{\odot}$, $r_{int,s}$ = 1.29 $\times$ 10$^{-6}$ pc and $r_{out,s}$ = 4.51 $\times$ 10$^{-5}$ pc for the secondary SMBH.}
    \label{double_sed}
\end{figure}

The SED for the primary SMBH emission is represented by a red solid line in the top panel of Fig.~\ref{double_sed} while the SED for the secondary SMBH emission is shown by the red solid line in the middle panel of Fig.~\ref{double_sed}. In both cases, the 100 emitting radii are shown in gray. The total SED resulting from the combined emission from the two accretion disks around the two SMBHs is shown by the red solid line in the bottom panel of Fig.~\ref{double_sed}. The first peak in the far-UV corresponds to the secondary SMBH accretion disk emission with a mass of $M_{*,s}$ = 4.5 $\times$ 10$^6 M_{\odot}$. The accretion disk around it emits in the range 0.1 $\mu$m to 0.3 $\mu$m. The second peak in the UV-optical corresponds to the primary SMBH emission with a mass of $M_{*,p}$ = 1.5 $\times$ 10$^{8} M_{\odot}$ and has an accretion disk that emits from 0.25 $\mu$m to 1 $\mu$m. This simulation of a binary SMBH system reproduces the cut-off observed in the near-UV flux spectra around 0.4 $\mu$m, which has been reported in previous studies \citep{Smith_1995, Leighly2014, Yan_2015}.

\subsection{Additional components}
\label{Modeling:others}

To construct a comprehensive SED for Mrk~231, it is essential to incorporate all significant emitting components of an AGN system. In this context, we added the infrared emission from the dusty torus surrounding the potential binary system of SMBHs in Mrk~231. For our model, we used the SED library of dusty torus simulations available provided by \citet{Siebenmorgen_2015}. It consists of a comprehensive library of SED models for AGNs with a two-phase torus structure around SMBHs. In this library, SEDs are computed for AGNs at a distance of 50 Mpc and a luminosity of 10$^{11} L_{\odot}$, which we adapted for Mrk~231. 

We also included in our model the optical emission from the host galaxy of Mrk~231. To do so, we used the survey established by \citet{Polleta} that contains a collection of 25 spectral models for various extragalactic objects such as elliptical galaxies at different ages, spirals, starbursts, and AGNs of types 1 and 2. These models cover a wide range of wavelengths, from 1000 \AA \ to 1000 $\mu$m, and were created using the \textsc{grasil} code. The host galaxy we used is a 13 Gyr old elliptical galaxy, similarly to what is observed for Mrk~231 \citep{Hamilton_1987}. 

\begin{figure}
    \centering
    \includegraphics[width=1\linewidth, trim={5cm 1.5cm 19cm 7cm}, clip]{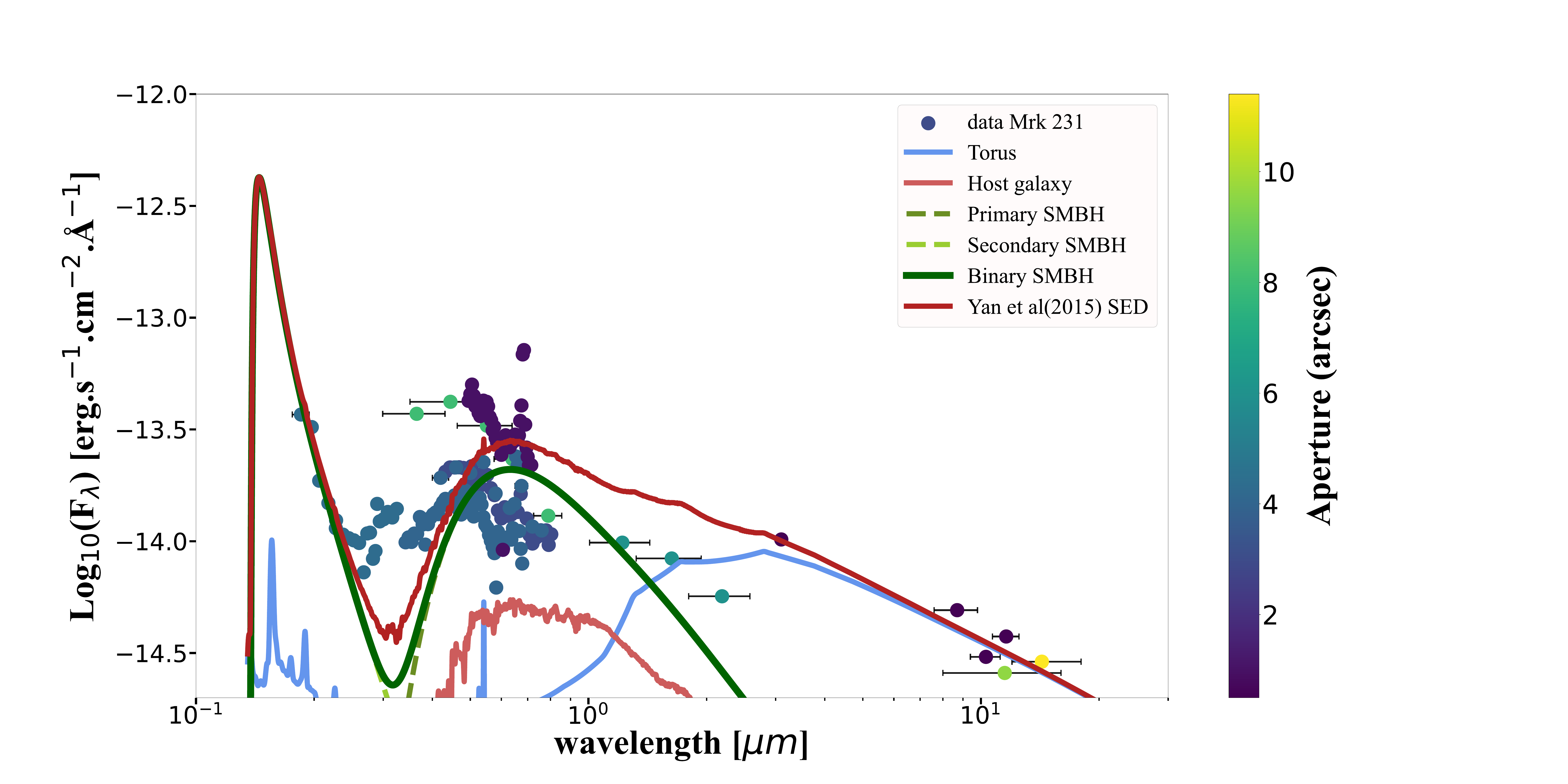}
    \caption{Logarithm of the total flux as function of the wavelength of Mrk~231. The dark green line corresponds to the modeled SED of the binary SMBH system, with the light green dotted lines denoting the secondary SMBH ($M_{*,s}$ = 4.5 $\times$ 10$^{6} M_{\odot}$) and the darker green dotted one is for the primary SMBH ($M_{*,p}$ = 1.5 $\times$ 10$^{8} M_{\odot}$), following \citet{Yan_2015}. The light red line is for the host galaxy and the blue one is for the torus component. The dark red line is the total SED with the following set of parameters : $M_{*,p} , M_{*,s} , \theta_p , \theta_s ,\alpha_{p} ,\alpha_{s}, p_{p} , p_{s}, a_{BBH}, f_{r,s}$.}
    \label{components_mrk231_}
\end{figure}

Figure~\ref{components_mrk231_} shows the superposition of each of the components of our model on top of the total flux data from our compilation. The two SMBH disk emission components are represented by green dotted lines, with the secondary SMBH displaying substantial emission in the UV spectrum, spanning from 0.1 to 0.3  $\mu$m, while the primary SMBH's emission extends from 0.3 to 1 $\mu$m. The sum of the two contributions is shown using a solid green line. The host galaxy flux is shown using the solid red line in Fig.~\ref{components_mrk231_}, with a pronounced starlight emission in the range 0.4 to 2 $\mu$m. Finally, the torus flux exhibits strong infrared emission across the wavelength range 1 to 11 $\mu$m. From this figure, it is evident that the binary SMBH model established using the parametrization from \citet{Yan_2015} does not instantaneously fit the observed data well. A fine-tuning of the parameters of the model is therefore necessary.

\subsection{Fitting the total flux data}
\label{Modeling:fit_flux}

As shown in Fig.~\ref{components_mrk231_}, the previous model does not properly reproduce the observations. Despite a reasonably accurate alignment with the data around 0.2 $\mu$m, which is attributed to the emission of the secondary SMBH, the modeled flux deprivation around 0.3 $\mu$m, which indicates the transition between the emissions of the two SMBHs, is significantly deeper and wider than what is observed. Moreover, the emission of the primary SMBH, spanning from 0.3 $\mu$m to 1 $\mu$m, appears to be inadequately aligned, requiring a shift to shorter wavelengths to improve the consistency with observations. These inconsistencies highlight limitations in the current parametrization of this model, particularly on the evaluation of the contribution of the two SMBHs. Adjustments to the underlying assumptions could lead to a better correspondence with our archived data. 

Therefore, we modified this model by fine-tuning the following set of parameters : $\theta \equiv (M_{*,p} , M_{*,s} , \theta_p , \theta_s ,\alpha_{p} ,\alpha_{s}, p_{p} , p_{s}, r_{int,p} , r_{out,p} , r_{int,s}, r_{out,s})$. Firstly, in order to adjust our simulation model, we investigated the influence of each parameter on the SED of the binary SMBH system. Some parameters can significantly affect the SED, such as SMBH mass, the semimajor axis, and the mean Roche radius. The aim here is simply to reproduce the model of the SMBH binary system; thus, we adjusted the parameters of our model manually, trying to reconcile the model with observed data. Secondly, we updated the radii of our SMBHs accretion disks, adding four new parameters to the binary SMBH model SED : the outer and inner radii of the primary and secondary SMBH ($r_{int,p} , r_{out,p} , r_{int,s}, r_{out,s}$). With this approach, where we incorporated four radii parameters; therefore, we had to remove two others : the semimajor axis, $a_{BBH}$, and the Roche lobe radius fraction, compared to the model established by \citet{Yan_2015}. By changing the values of our accretion disks radii, the SMBH emission can be moved to different wavelengths and allow our model to reproduce the crucial flux deprivation around 0.3 $\mu$m. Afterwards, we could recalculate the value of $a_{BBH}$ and the Roche lobe radius fraction values using Eqs..~\ref{a_bbh} and \ref{f_r_s}, with $q$ the SMBH mass ratio :

\begin{equation}
    a_{BBH} = R_{int,p}\frac{(q+1)}{(q+1)\left(\frac{M_{*,s}}{3M_{*,p}}\right)^{\frac{1}{3}}+1},
    \label{a_bbh}
\end{equation}

and

\begin{equation}
    f_{r,s} = R_{out,s} \frac{0,6 \ q^{\frac{2}{3}} + log(1+q^{\frac{1}{3}})}{(0,49 \ a_{BBH} \ q^{\frac{2}{3}})}.
    \label{f_r_s}
\end{equation}

To obtain the best fit of our models to Mrk~231 data and constrain our model parameters, we used a $\chi^{2}$ minimization. Our binary SMBH SED model includes 12 free parameters: $\theta \equiv (M_{*,p} , M_{*,s} , \theta_p , \theta_s ,\alpha_{p} ,\alpha_{s}, p_{p} , p_{s}, r_{int,p} , r_{out,p} , r_{int,s}, r_{out,s})$,  which were optimized through a $\chi^{2}$ minimization process. Our best fit between the model and the observational data is shown by the red curve in the top-left panel of Fig.~\ref{fit_data_mrk231}. The $\chi^{2}$ minimization allowed us to determine the optimal parameter values, ensuring the best agreement between our model and the data. During the minimization process for the total SED model, we strictly focused on the emission from the binary SMBH system, which extends over the wavelength range from 0.1 to 1 $\mu$m. The host galaxy and dusty torus emissions have been kept fixed. In addition, compared to the initial total SED model presented in Fig.~\ref{components_mrk231_}, we applied a slight normalization to the host galaxy emission. This adjustment increased the contribution of the host galaxy in the wavelength range from $\sim$ 0.35 to 2 $\mu$m. The final, best-fit parameters for the binary SMBH system are $M_{*,p} = 1.62 \times 10^{8} M_{\odot} , M_{*,s} = 11.00 \times 10^{6} M_{\odot} , \theta_{p} = 35.26^{\circ} , \theta_{s} = 43.83^{\circ}  ,\alpha_{p} = 0.07 ,\alpha_{s}= 0.14, p_{p}=0.74 , p_{s}= 0.08, r_{int,p} = 0.86 \times 10^{-3} pc , r_{out,p} = 1.33 \times 10^{-3} pc, r_{int,s} = 1.97 \times 10^{-6} pc, r_{out,s} = 4.20 \times 10^{-4} pc$. All parameters values are summarized in Table~\ref{param_flux}. Based on these results, we determined the value of the semimajor axis and the Roche lobe factor using Eqs.~\ref{a_bbh} and \ref{f_r_s}. We found $a_{BBH} = 146 $ AU and $f_{r,s} = 3.2$.

\begin{table}
    \centering
    \caption{Parameters and best-fit values for the total flux model.}
    \begin{tabular}{ c c c } 
        \hline
        Parameter & Best-fit value & Units \\ 
        \hline 
        $M_{*,p}$ & (1.62 $\pm$ 0.10) & 10$^{8} M_{\odot}$\\
        $M_{*,s}$ & (1.1 $\pm$ 5.5 $\times$ 10$^{-7}$) & 10$^{7} M_{\odot}$ \\
        $\theta_p$ & (35.26 $\pm$ 6.99) & $^{\circ} $ \\
        $\theta_s$ & (43.83 $\pm$ 6.29) & $^{\circ} $  \\
        $\alpha_{p}$ & (0.076 $\pm$ 0.003) & - \\
        $\alpha_{s}$ & (0.144 $\pm$ 0.007) & - \\
        $p_{p}$ &  (0.741 $\pm$ 0.003) & - \\
        $p_{s}$ &  (0.079 $\pm$ 0.003) & - \\
        $r_{int,p}$ & (2.66 $\pm$ 0.05) & 10$^{15}$ [cm]\\
        $r_{out,p}$ & (4.12 $\pm$  0.10) & 10$^{15}$ [cm]\\
        $r_{int,s}$ & (0.0060 $\pm$ 0.0002) & 10$^{15}$ [cm]\\
        $r_{out,s}$ & (1.297 $\pm$ 0.007) & 10$^{15}$ [cm]\\
        \hline 
    \end{tabular}
    \label{param_flux}
\end{table}

Some of our findings are aligned with those reported by \citet{Yan_2015}, especially regarding several parameters of the primary SMBH, such as its mass ($M_{*,p}$ = 1.62 $\times$ 10$^{8} M_{\odot}$) and its inclination angle ($cos (\theta_{p}) = 0.8$). However, we did not find the same Eddington ratio ($f_{Edd}$) as found by \citet{Yan_2015} for the primary SMBH; namely, they found $f_{Edd,p}$ = 0.5, calculated using $f_{Edd} = \dot{M}_{*,p}/\dot{M}_{Edd}(M_{*,p})$, with $\dot{M}_{*,p}$ as the accretion rate of the primary SMBH (as defined in Eq.~\ref{critical_accretion_rate}). In constrast, we found an Eddington ratio of $f_{Edd,p}$ = 0.057. Moreover, significant differences were observed for several parameters of the secondary SMBH. We determined a mass of $M_{*,s}$ = 1.1 $\times$ 10$^{7} M_{\odot}$ ; whereas \citet{Yan_2015} reported a value of 4.5 $\times$ 10$^{6}M_{\odot}$. Moreover, we found a value of $f_{Edd,s}$ = 0.011, in contrast with their reported ratio of 0.6 for the secondary SMBH (see below for the implications). Regarding the semimajor axis value, we obtained a value ($a_{BBH}$ = 146 AU) slightly lower than theirs ($a_{BBH}$ = 590 AU). Finally, our calculation of the mean Roche radius yielded to $f_{r,s}$ = 3.2, which is to be compared to their value of $f_{r,s}$ = 0.11. Regarding the last four parameters, we found a significantly different values for the outer radius of the primary black hole, likely due to the fact that the authors did not use the Toomre parameter as we did (instead they simply used $r_{out,p}$ = 10$^{5} r_{int,p}$). For the inner radius of the secondary SMBH, \citet{Yan_2015} imposed $r_{int,s}$ = 3.5 $GM_{*,s}/c^{2}$ ; whereas in our analysis, we used  $r_{int,s}$ = 3$\times$ ${2GM_{*,s}}/{c^{2}}$. We recalculated the value of each of the four radii using the parametization in \citet{Yan_2015} and found $r_{int,p}$ = 3.38 $\times$ 10$^{-3}$ pc, $r_{out,p}$ = 337.98 pc for the primary SMBH and $r_{int,s}$ = 0.75 $\times$ 10$^{-6}$ pc, $r_{out,s}$ = 0.68 $\times$ 10$^{-4}$ pc.

Overall, we found Eddington ratios ($f_{edd}$) that are significantly different from those determined by \citet{Yan_2015}. Such a ratio is used to compare the accretion rate of a system to its maximum possible value, with a value close to 1 indicating a high accretion rate and increased luminosity. \citet{Yan_2015} imposed an Eddington fraction of between 0.1 and 1, but recent studies have shown that the Eddington ratio can be considerably lower, of the order of 10$^{-2}$ and 10$^{-3}$ for many objects \citep{Raimundo}, which is in agreement with our best-fit parameters. Our results therefore suggest less accreting SMBHs than those estimated by \citet{Yan_2015}. Such finding makes the formation of a stable, luminous mini-disk around the primary SMBH even less likely than in \citet{Yan_2015}, which is the reason why such an additional component was neglected in both their analysis and ours. Moreover, we determine that the second SMBH is more massive than the one estimated by the authors. We also find a much smaller semimajor axis, indicating that the two SMBHs are closer than estimated by \citet{Yan_2015}. On the other hand, we obtained a larger mean Roche radius, which means that the limit at which the second SMBH would be deformed by tidal forces is higher than that estimated by \citet{Yan_2015}.

\subsection{Fitting the polarized flux data}
\label{Modeling:fit_pol}

In the second part of our study, which is the core of this analysis and a first approach in the field of photopolarimetric fittings of binary SMBH models, we analyzed the response of our model in polarization. We did not add any further variations to the geometrical parameters of the modeled system. 

To do so, for each component of the model, we associated a wavelength-independent polarized component (I, Q, U) based on the Stokes formalism. The Stokes formalism uses of a four-dimensional vector $\overrightarrow{S}$ representation described by four measurable Stokes fluxes \citep{Stokes_1852}. The polarization vector $\overrightarrow{S}$ is represented in Eq.~\ref{stokes_parameter_matrix}. The parameter I represents the total intensity of the radiation. The wave's linear polarization is characterized by Q and U, where Q is the difference between the vertical and horizontal polarization state ; U represents the difference between the linear polarization with orientations of +45$^\circ$ and -45$^{\circ}$ ; and V represents the state of circular polarization and stands as the difference between the right and left rotational direction (not used in this paper due to the lack of any circular polarization measurement for Mrk~231). Here, each component of $\overrightarrow{S}$ is expressed as a fraction of the total intensity, $I$. The parameters $P_1, P_2, P_3$, ranging from -1 to 1, represent the fraction of total intensity contributing to the polarized light :

\begin{equation}
\overrightarrow{S} = \begin{pmatrix} I \\ Q \\ U \\ V \end{pmatrix} = \begin{pmatrix} I \\ I . P_1 \\ I . P_2 \\ I . P_3 \end{pmatrix}.
\label{stokes_parameter_matrix}
\end{equation}

As Eq.~\ref{stokes_parameter_matrix} illustrates, all Stokes parameters can be expressed in terms of the total intensity, $I$, and a fractional parameter, which varies between -1 and 1. Specifically, $Q_{i}=q_{i} \times I_{i}$ and $U_{i} = u_{i} \times I_{i}$ where $i$ denotes the components of the AGN, including the primary and secondary SMBH, the host galaxy and the dusty torus. In the following analysis, the parameters $I_i$ correspond to the flux $F_{\lambda,i}$ for the best fit value of each element of the AGN. The ultimate goal is to determine the values of the parameters $u_i$ and $q_i$ for the four components of the AGN that reproduce best the polarization degree and position angle of Mrk~231. Indeed, these scalar parameters can be used to compute the polarization degree, $P$, and the polarization angle, $ \theta$, which are obtained from the Stokes parameters and defined in Eqs.~\ref{polarization_degree} and \ref{polarization_angle}, respectively :

\begin{equation}
    P = \frac{I_{pol}}{I} = \frac{\sqrt{Q^{2} + U^{2} + V^{2} } }{I},
    \label{polarization_degree}
\end{equation}

\begin{equation}
    \theta = \frac{1}{2} arctan\left(\frac{U}{Q}\right).
    \label{polarization_angle}
\end{equation}

The polarization degree is calculated as the ratio between the intensity of the polarized light and the total intensity (Eq.~\ref{polarization_degree}) and the polarization angle is calculated using Eq.~\ref{polarization_angle}, with $arctan$ denoting the quadrant-preserving inverse of the tangent function. If $P = 0$, the wave is said to be unpolarized and if $P=1$, the wave is said to be fully polarized. If $0<P<1$, the wave is said to be partially polarized. These are the values we can compare to observations.

Based on the established minimized SED model of Mrk~231, we determined the best-fit polarization degree, polarization angle, and polarized flux from the wavelength-independent\footnote{In order to align our model more closely with observed data, which shows a slight rotational variation in the polarization angle from approximately 105$^{\circ}$ to 130$^{\circ}$ over the wavelength range of 8 to 20 $\mu$m, we had to introduce a linear component described by the equation m$\lambda$ + c with $m$ calculated as $m$ = (136-105)/(20-0.1) and $c$ being determined by the relation $c$ = 105$-m \times$8.1. Such wavelength-dependence is expected for dust scattering, absorption and emission processes.} I, Q, and U Stokes parameters. Our polarized model is thus composed of six free parameters : $(q_{*,p}, q_{*,s}, q_{torus}$, $u_{*,p}, u_{*,s}, u_{torus}$. Since the host galaxy exhibits very low integrated polarization (generally well below 1\%), we set $U_{galaxy} = Q_{galaxy} = 0$ \citep{Simmons_2000}. Again, we used a $\chi^{2}$ minimization to find the optimized parameter values for the degree and angle of polarization independently. Because Stokes parameters are scalar, we averaged the best-fit values from the three independent (but very similar) best-fit parametrizations in the polarized flux, polarization degree, and polarization angle to obtain the value of $q_{i}$ and $u_{i}$ for each element that constitutes the AGN (primary and secondary SMBH, host galaxy, and dusty torus). We found $q_{*,p}$ = -0.2254, $q_{*,s}$ = -0.0012, $q_{torus}$ = -0.0066, $u_{*,p}$ = 0.0047 , $u_{*,s}$ = -0.0310, and $u_{torus}$ = -0.0064. The host galaxy polarization was set to zero and our minimization process validated this hypothesis. 

\begin{figure*}
    \centering
    \includegraphics[width=1\linewidth, trim={18cm 10cm 19cm 13cm}, clip]{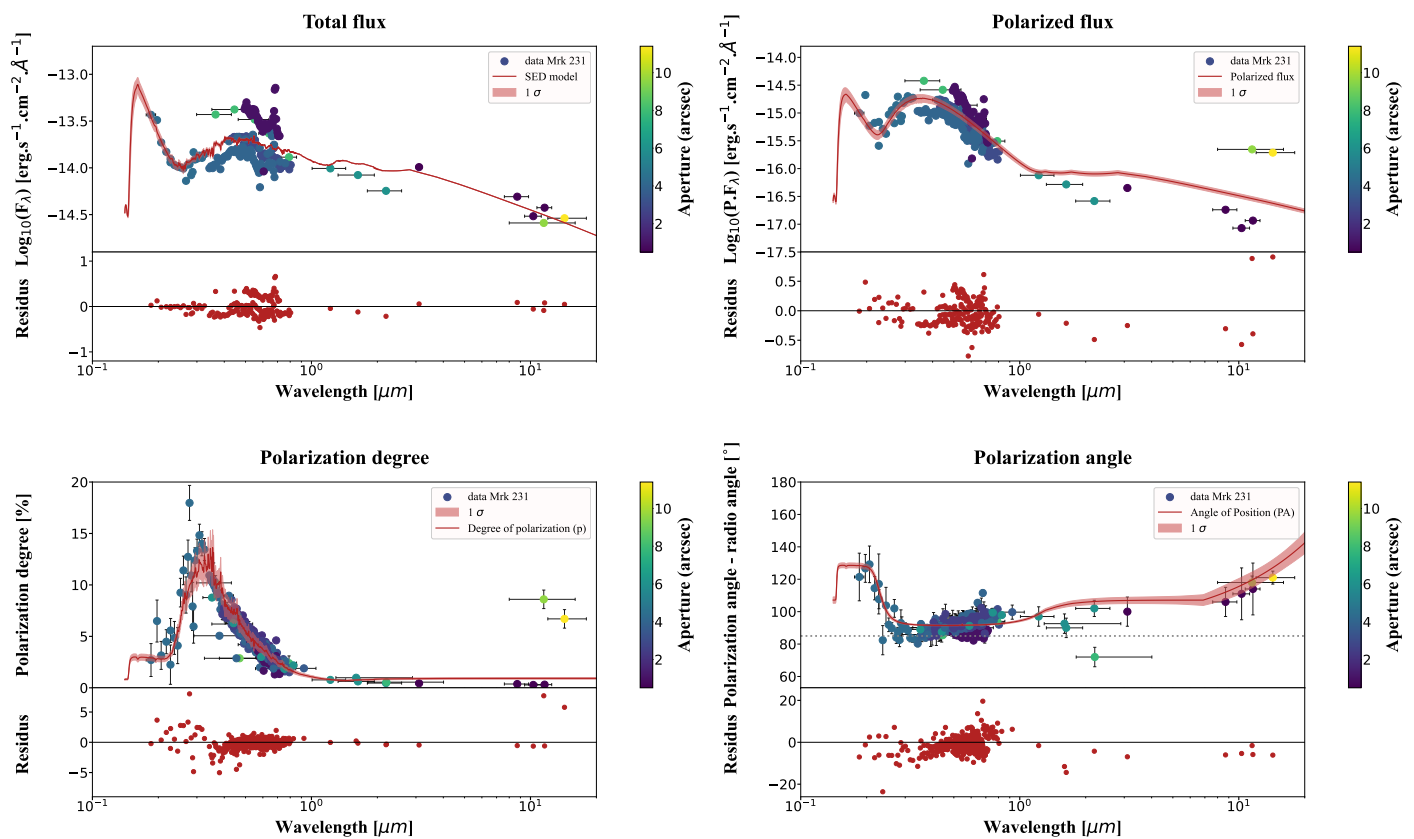}
    \caption{\textit{Top left panel} : Mrk~231 total flux as a function of wavelength. \textit{Top right panel} : Polarized flux as a function of wavelength, established from the multiplication of the flux and the polarization degree. \textit{Bottom left panel} : Degree of polarization as a function of wavelength. \textit{Bottom right panel} : Polarization position angle minus the radio position angle (5$^{\circ}$, see \citealt{Ulvestad_1999}) as a function of wavelength. Below each panel, we give the residuals between the observational data and the theoretical model, plotted across the wavelength range.}
    \label{fit_data_mrk231}.
\end{figure*}

Our best-fit parametrization is represented by the solid red line in the top-right (polarized flux), bottom-left (polarization degree), and bottom-right (polarization angle) panels of Fig.~\ref{fit_data_mrk231}. All our $q_{i}$ and $u_{i}$ values are reported in Table~\ref{param_pol_final}. The degree of polarization for each element of the AGN is $P_p$ = 23.04 $\pm$ 2.76 \% for the primary black hole, $P_s$ = 3.10 $\pm 0.36$ \% for the secondary black hole, and $P_{torus}$ = 0.92 $\pm$ 0.08 \% for the dusty torus. We calculated the corresponding polarization position angle for each element that contributes the observed polarization properties of the AGN and we found $\theta_p$ = 95.96$^{\circ}$ $\pm$ 0.17$^{\circ}$ for the primary SMBH, $\theta_s$ = 133.85$^{\circ}$ $\pm$ 1.56$^{\circ}$ for the secondary SMBH, and $\theta_{torus}$ = 112.14$^{\circ}$ $\pm$ 2.04$^{\circ}$ for the torus component.

The binary SMBH model suggested by \citet{Yan_2015} and modified in this paper does not only correctly fits the total flux SED but is remarkably well aligned with the observed polarization signatures of Mrk~231. We recall that the total flux, the polarization degree and the polarization angle are three independent observable, so fitting all three of them at once with the same model provides a strong support for a potential binary SMBH at the earth of Mrk~231. The observed polarization degree and angle coincide strongly with the curves from our analytical model but something even more important should be noted regarding the polarized flux. The binary SMBH model predicts not one but two bumps in polarized flux. The first one is in the far-UV while the other one is in the optical. The optical bump is clearly seen in the polarized data, but their is a lack of measurements in the far-UV to definitively prove the presence of a secondary, less-massive SMBH. This is a strong prediction that could be potentially examined in the future if a space-based, far-UV polarimeter is pointed towards this source.

\begin{table}
    \centering
    \caption{Parameters and best-fit values for the polarization model.}
    \begin{tabular}{ c c c }
        \hline
        Parameter & Value \\ 
        \hline
        $q_{*,p}$ & -0.2254 $\pm$ 0.0282 \\
        $q_{*,s}$ & -0.0012 $\pm$ 0.0016 \\
        $q_{*,torus}$ & -0.0066 $\pm$ 0.0009 \\
        $u_{*,p}$ & -0.0475 $\pm$ 0.0009 \\
        $u_{*,s}$ & -0.0310 $\pm$ 0.0036 \\
        $u_{*,torus}$ & -0.0064 $\pm$ 0.0003 \\
        \hline
    \end{tabular}
    \label{param_pol_final}
\end{table}


\section{Radiative transfer modeling}
\label{STOKES}

In the previous section, we describe how the peculiar wavelength-dependent, far-UV to near-IR polarization degree and angle of Mrk~231 can be explained by a binary SMBH system if the large one (producing the optical/near-IR spectrum thanks to an extended yet truncated accretion disk) is responsible for an observed polarization degree of $\sim$ 23\% at $\sim$  96$^\circ$ and if the smaller one (orbiting the first black hole and whose disk is responsible for the UV polarization) is at the origin of an observed polarization degree of $\sim$ 3\% at $\sim$ 134$^\circ$. 

\subsection{Polarization from the SMBH companion}
\label{STOKES:UV}

Producing a polarization of about 3\% is not complicated in the framework of thermal emission and scattering within an accretion flow, particularly if it is highly ionized. \citet{Chandrasekhar1960} has shown that the polarization degree emerging from a disk dominated by Thomson scattering with an infinite optical depth is inclination-dependent and can be as high as 11.7\% in certain cases. This, of course, depends on the exact geometry of the scattering medium and the vertical and horizontal optical depths of the accretion structure \citep{Sunyaev1985,Webb1986,Phillips1986,Kartje1991}. We ran the Monte Carlo radiative transfer code {\sc stokes} \citep{Goosmann2007,Marin2012,Marin2015,Marin2018,Rojas2018} to determine what is the inclination range able to reproduce our estimated polarization degree from the secondary SMBH, considering a simple disk geometry illuminated from the base of the accretion structure with arbitrary initial polarization vectors and directions of propagation. The disk is filled with electrons to reach the Thomson limit. 

\begin{figure}
    \centering
    \includegraphics[width=1\linewidth, trim={0cm 0cm 0cm 0cm}, clip]{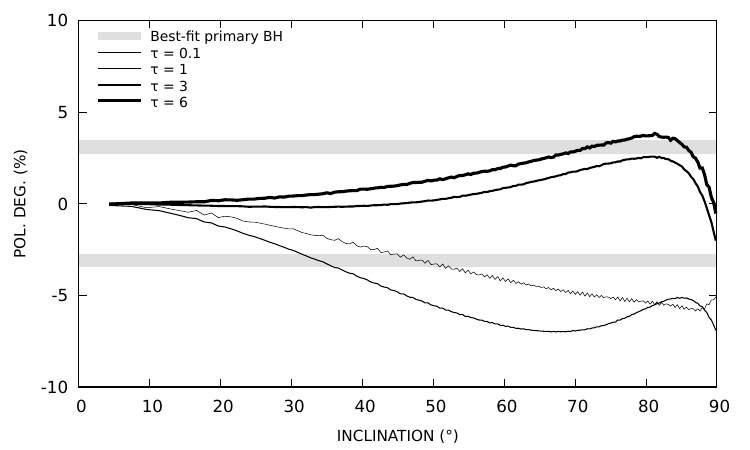}
    \caption{Optical polarization emerging from an accretion disk filled with electrons as a function of the viewer's inclination.  Different vertical optical depth are investigated. The wavelength-independent polarization degree emerging from the secondary SMBH disk estimated from our analysis is reported in gray. By convention, polarization vectors oriented parallel to the projection of the disk symmetry axis on the plane of the sky were assigned a negative value.}
    \label{Fig:MC_disk}
\end{figure}

Results are shown in Fig.~\ref{Fig:MC_disk} and nicely reproduce similar figures reported in the literature, for instance, Fig.~1 in \citet{Kartje1991}. We see that a typical disk-like accretion disk can already produce a polarization degree of 3\% if the inclination of the system is about 70$^\circ$ (from the vertical axis of the system) for optically thick disks. In the case of marginally thick disks ($\tau_e$ = 1), the disk only has to sustain an inclination of 30-35$^\circ$. On the other hand, in the optically thin case ($\tau_e$ = 0.1), the derived inclination is 44-51$^\circ$, which coincides with the inclination angle of the disk obtained from our fits (see Table~\ref{param_flux}). The polarization we observe may then originate from the secondary accretion disk atmosphere, rather from its internal parts. The fact that the polarization angle resulting from those simulations is either parallel or perpendicular to the symmetry axis of the system, and that the observed UV polarization angle is about 30$^\circ$ different than perpendicularity tells us that the secondary SMBH's disk is not coplanar to the most massive one. It orbit is inclined in comparison with the symmetry plane of the most massive SMBH.

\subsection{Polarization from the primary SMBH}
\label{STOKES:OPT}

Now, the difficulty is explaining how to produce a polarization as high as 23\% and perpendicular to the AGN radio axis. Scattering from a disk cannot achieve such high polarization levels, as previously demonstrated in Fig.~\ref{Fig:MC_disk} and by \citet{Chandrasekhar1960}. However, taking into account the fact that Mrk~231 is a BAL quasar, meaning that the line of sight of the observer crosses a large column of ejected material in the form of a wind, we can investigate whether the polarization from the primary black hole component is not simply due to scattering of the thermally emitted photons inside disk winds. 

\begin{figure}
    \centering
    \includegraphics[width=1\linewidth, trim={0cm 0cm 0cm 0cm}, clip]{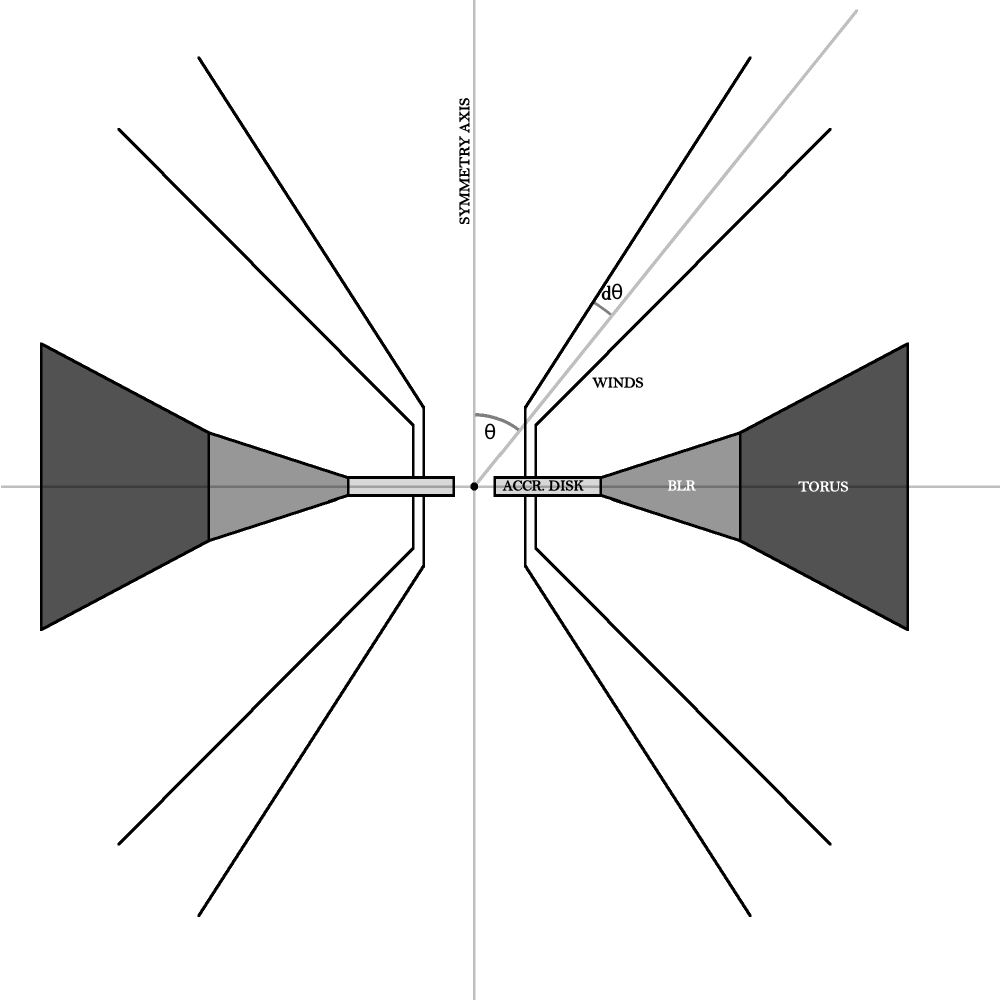}
    \caption{Schematic view of the model used in Sect.~\ref{STOKES:OPT} to reproduce the high, 23\% continuum linear polarization degree attributed to the central SMBH. The image is not to scale and only meant to represent the geometry of the various components of the system. It includes a central SMBH and a disk-like accretion structure from which emerges powerful winds through radiation and line pressure; $\theta$ represents the outflowing angle of the wind and d$\theta$ its half-opening angle. Along the equatorial plane are also the region responsible for the emission of the broad lines (the BLR) and the torus. See the text for details and the geometrical parametrization of the components.}
    \label{Fig:Model}
\end{figure}

Such a hypothesis was already mentioned by a number of observers, who found that the polarization of BAL quasars is often high (of a few percent) and almost always perpendicular to the radio jet axis of the object \citep{Moore1984,Ogle1999,Lamy2004,DiPompeo2013}. We ran the radiative transfer code {\sc stokes} once again but considering a disk wind structure, as already explored in \citet{Marin_2013}, based on the work of \citet{Elvis2000}. In this scenario, see Fig.~\ref{Fig:Model}, the accretion disk is the seat of a large ejection of matter in the polar axis of the AGN, because of the radiation and line pressures exerted by the emission of the inner edges of the disk \citep{Proga1998,Proga2000,Proga2004}. The outer edge of the accretion disk is less impacted and the broad line region (BLR) and the torus can form, maybe in the form of a failed accretion disk wind \citep{Czerny2011}. 

We included all those regions into our three-dimensional model, with the following constraints : the disk emits isotropic, unpolarized radiation between 1000 and 10\,000~\AA\,  within the accretion disk itself. The vertical, Thomson optical thickness of the disk has been set to 3 and it extends up to 1~mpc, as suggested by our previous analysis. The end of the accretion structure is where the BLR onsets, which has a half-opening angle set to 20$^\circ$ with respect to the equatorial plane. This value was chosen in order to match the flattened geometrical configuration of the BLR as expected from simulations \citep[see, e.g.,][]{Piotrovich2015}. The end of the BLR coincides with the inner wall of the circumnuclear torus (0.1~pc), parametrized here as a flared disk (similarly to the BLR), with a larger half-opening angle (40--45$^\circ$). This values corresponds to the expected half-opening angle of AGN tori, as reported by \citet{Shen2010}, \citet{Sazonov2015}, and \citet{Marin2014,Marin2016}. The torus extends up to 5~pc, but the outer radius as little impact onto the resulting polarization in the optical band. Most importantly, an accretion disk wind onsets at about a third of the accretion disk radial length and extends in a direction fixed by an user-selected value $\theta$. The half-opening angle d$\theta$ of the wind was fixed to 5$^\circ$ in this case. The winds are either filled with electrons or dust (whose composition is that of the Milky Way, see \citealt{Goosmann2007}) at different optical depths : 1, 2.5, and 5. Indeed, observations have proven that BAL QSO winds are significantly Thomson-thick along the radial flow direction \citealt{Elvis2000}.

\begin{figure}
    \centering
    \includegraphics[width=1\linewidth, trim={0cm 0cm 0cm 0cm}, clip]{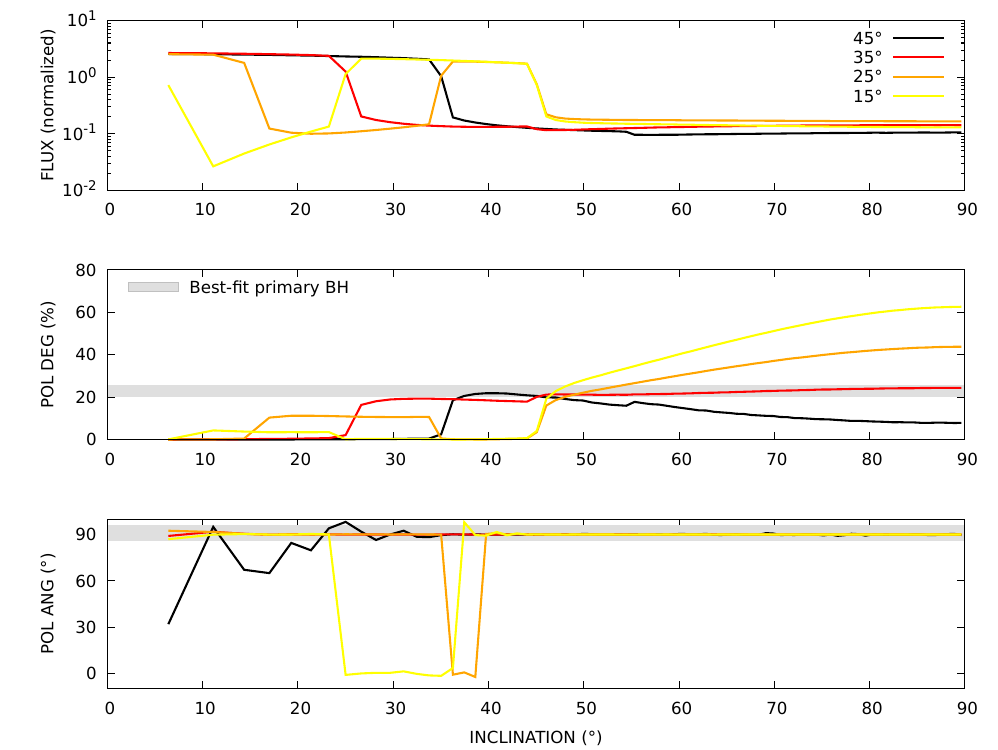}
    \caption{Total flux (normalized to the peak of emission, top), continuum linear polarization degree (middle) and polarization angle (bottom) as a function of the observer's viewing angle, defined with respect to the symmetry axis of the model, see Fig.~\ref{Fig:Model}. Four different inclinations of the wind outflowing direction are shown : 15$^\circ$, 25$^\circ$, 35$^\circ$, and 45$^\circ$. The wind is fully ionized. In gray is the polarization expected from the primary component of the SMBH binary found by our best-fit analysis.}
    \label{Fig:Simulation_best}
\end{figure}

We present in Fig.~\ref{Fig:Simulation_best} the wavelength-integrated simulation that best reproduces the degree and angle of polarization expected for the primary component of the SMBH binary. All other radiative transfer results are shown in the appendices for completeness. A fully ionized wind, inclined by 45$^\circ$ with respect to the symmetry axis of the system (the jet direction), with an half-opening angle of 5$^\circ$ and an optical thickness of 5 is able to produce a polarization degree that matches the polarization level expected from our spectral decomposition. The associated polarization position angle is 90$^\circ$, namely, it is perpendicular to the jet direction, as expected from observations. Such polarization is only reachable if the observer's inclination matches the orientation of the winds, which allows us to constrain the overall tilt of the system at the same time. 

~\

~\

Ultimately, our simulations proved that it is entirely possible to produce a polarization of 3\% by scattering in an accretion disk to explain the degree of polarization of the secondary black hole (the smaller of the two, whose accretion disk is responsible for the peak of the ultraviolet emission) and a polarization of 23\% by electron scattering in the quasar winds, associated with the primary black hole (the more massive, whose accretion disk is responsible for the optical emission). Additionally, the fact that electron scattering is favored against dust scattering also corroborates our hypothesis that the underlying polarization of the SMBH components are wavelength-independent, strengthening our analysis and our results.


\section{Discussion}
\label{Discussion}

\subsection{A picture of the system}
\label{Discussion:picture}

\begin{figure*}
    \centering
    \includegraphics[width=1\linewidth, trim={3.5cm 2cm 1cm 2cm}, clip]{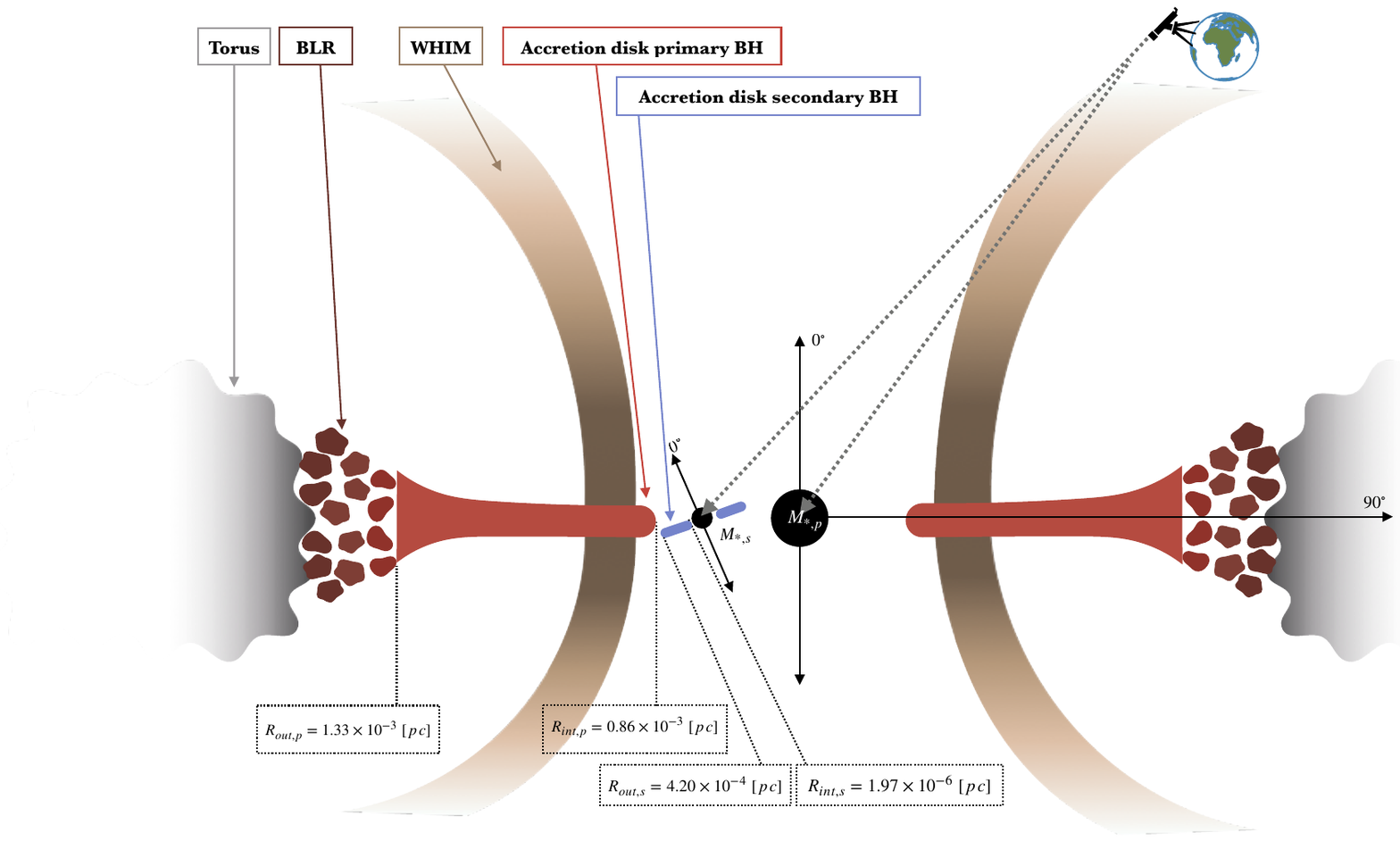}
    \caption{Schema of the structure of Mrk~231, no to scale, based on the model of the warm highly polarized matter (WHIM) and on our own conclusions obtained during this study.}
    \label{structure_mrk231}.
\end{figure*}

Our analytical model, supported by radiative transfer simulations, allows us to draw a picture of the unresolved core of Mrk~231. We present it as a cartoon in Fig.~\ref{structure_mrk231}. It contains all the information we deduced from our analysis. At the center of the (unscaled) sketch is the primary SMBH, within which orbits a secondary, lighter SMBH. This secondary SMBH is surrounded by a small accretion disk that only emits in the UV, since its outer radius corresponds to the inner radius of the accretion structure of the primary SMBH. The two are not coplanar, has revealed by the different polarization angles found by our spectral decomposition. The disk is probably oriented with an axis of symmetry inclined 130$^\circ$ (observed polarization angle) - 90$^\circ$ (expected polarization angle) = 40$^\circ$ relative to the north. The disk is likely seen at an angle of 70$^\circ$ relative to the observer's line-of-sight (see previous section). The primary SMBH and its surrounding components (its accretion disk, the disk-born winds, the BLR, and the torus) are though to be seen with in inclination of 40-50$^\circ$, with the observer's line-of-sight passing through the out-flowing material. This is also in agreement with the estimated inclination angle of BAL quasars \citep{Turnshek1988}.

\subsection{Arguments against a binary SMBH}
\label{Discussion:against}

The central engine of Mrk~231 has been the subject of numerous studies over the years, giving way to a variety of hypotheses, each based on specific analysis of the many particularities of this object. In this paper, we simulated the SED of Mrk~231 based on the model of a binary SMBH proposed by \citet{Yan_2015}. Following this model, the lowest-mass SMBH is surrounded by a mini accretion disk emitting in the far-UV and is located inside the accretion disk of the highest-mass SMBH, which emits in the near-UV and optical bands. The resulting depression observed in the spectrum of Mrk~231, around $\sim$ 3000 \AA, is attributed to the particular geometry of the binary system, which creates a gap between the emission regions. However, several studies have challenged this model and argue against the potential presence of a binary SMBH system in Mrk~231.

\citet{Veilleux_2016} observed a remarkable stability in Mrk~231 continuum emission over a three-year period between 2011 and 2014, with no significant variation. According to the authors, if the mini-disk is responsible for the UV emission, periodic flux variations on the scale of the orbital period would be expected ($t_{orb} \approx$ 1.2 years \citealt{Yan_2015}) ; however, this has not been observed. Nonetheless, they noted a slight variation around $\sim$1300 \AA \ between archival International Ultraviolet Explorer and their observations with the Cosmic Origins Spectrometer (COS) from the HST. According to their analysis, the absence of periodic variations in the UV/optical spectrum would imply the absence of a binary SMBH in the core of Mrk~231. However, this argument was challenged by recent studies who identified a slight periodic variability in optical photometric data from Catalina Surveys Data Release 2, which could suggest a periodic variability in the optical emission associated with a binary SMBH system \citep{Kova_2020}. The authors used Lomb–Scargle periodogram and 2DHybrid method to detect a period of $\sim$ 1.1 year with a confidence level of 99.7 \%, which should correspond to the orbital period of the binary system and is in agreement with $t_{orb}\sim$ 1.2 year. This is corroborated by \citet{Yang_2018}, who recently detected significant UV variability with Swift, in agreement with the UV emission of an accretion disk as proposed by the binary SMBH model. Since the two accretion disks are not coplanar, we expected a periodic variation in the flux and polarization on a timescale comparable to the orbital separation of the binary SMBH system. A dedicated monitoring study in polarization, particularly in the far-UV waveband, would be essential to track these variations, as this spectral region corresponds to the secondary SMBH emission. Future observations with instrument capable of high-sensitivity measurement of far-UV polarization could provide valuable constrains on the disk precession and the geometry of the intern structure of Mrk 231.

\citet{Veilleux_2016} also reported a stable Ly$\alpha$ profile over the same three-year period (2011 - 2014), using Cycle 19 and Cycle 21 HST/COS data. According to the binary model proposed, the mini disk of the secondary SMBH emitting in the far-UV should dominate the line emission in the far-UV and broad blueshifted Ly$\alpha$ profile should be observed over the $\sim$ 1.2~yr orbital period, but it has not been detected. Moreover, absorption lines are only detected on the near-UV/optical waveband. This implies that the broad absorption line region is found only in the direction of the near-UV/optical emitting region - and not in the direction of the far-UV mini disk, strengthening the idea that the two sources of emission are distinct (see our schematic view of the AGN). According to \citet{Veilleux_2016}, within the context of the binary model, we should expect velocity differences between the far-UV and near-UV/optical emission lines, but it has not been detected either. To challenge the binary SMBH model, \citet{Leighly2016} analyzed several emission lines of Mrk~231, but only in the near-IR : He I$^{*}\lambda$10830, P$\beta$ $\lambda$12818 \AA, P$\alpha$18751 \AA. If the BLR emission is the result of photoionization produced by the far-UV mini disk, \citet{Leighly2016} would have detected large equivalent width (about $\sim$ 100 times larger than normal); however, the extrapolated IR photoionizing continuum is 100 times weaker than the observed continuum using the binary model proposed by \citet{Yan_2015}, so no such broad lines were found. Another prediction of the binary model is that because the Fe II optical emission is very strong, Mrk~231 should also have strong UV Fe II emission but it as not been observed. \citet{Leighly_2014} showed that this lack of UV Fe II emission is caused by reddening, while \citet{Yan_2015} proposed that the reddening is minimal and suggested that this lack of UV Fe II emission is due to Mrk~231's classification as FeLoBAL (Fe II emission being absorbed by the BAL outflow). \citet{Leighly2016} disagreed with this assessment because the velocity of the low-ionization absorption line velocity shift is between around -5500 and -4000 km.s$^{-1}$; whereas, it would take between -8000 and -10000 km.s$^{-1}$ to no longer have Fe II UV emission according to the binary model proposed. Anyway, it has been shown previously that the far-UV emission is little affected by the BAL wind, so absorption of the Fe II emission is unlikely. The reason why the equivalent widths of the BLR induced emission line is not as high as expected by \citet{Leighly2016} likely lies in the fact that the far-UV mini disk is much less accreting that what was suggested by \citet{Yan_2015}. This is because our analysis determined an Eddington ratio of this component, that was a factor of 10 lower than the prescription detailed in \citet{Yan_2015}.

Past polarimetric studies have examined the high polarization of 15$\%$ at $\sim$ 3100 \AA \ that subsequently decreases between 3100 to 1600 \AA \ \citep{Smith_1995} to understand what is going on in the core of Mrk~231. This decline in polarization have been explained by \citet{Wills_1992} by invoking electron scattering producing a blue polarized spectrum; however, since dust scattering is strongly wavelength-dependent \citep{Smith_1995}, \citet{Leighly2016} favored scattering on dust grains in a large and diffuse region rather than from a compact source associated with the far-UV mini accretion disk. This decreasing trend in the UV has also been explained as dilution by emission of obscured and unpolarized stellar light or by a binary SMBH whose far-UV emission would be unobscured and slightly polarized, while the primary component emitting in the optical would be obscured and strongly polarized \citep{Veilleux_2016, Leighly2016,Leighly_2014}. This last scenario is clearly favored by our study.

The UV continuum deficit hypothesis as a signature of a binary SMBH was also criticized by \citet{Leighly2016}. They suggested instead that it could result from reddening plus a far-UV contribution from starburst activity in the host galaxy. In addition, the authors demonstrated that the continuum UV emission did not emanate from a compact center region, but rather from a wider region, which could not be subject to rapid variability. This is hardly compatible with simulations of the pan-SED of starburst galaxies presented by \citet{Groves2008}, for instance, since starbursts cannot produce strong far-UV peaks of emission; however, the search for similar SED in a catalog of 138 binary SMBH candidates by \citet{Guo2020} was inconclusive. Only six systems presented the same abnormally red, blue-truncated spectra but this fraction is consistent with the fraction of similar outliers in a control sample that could very well be explained via dust-reddening. 

It conclusion, the question is far from being completely resolved, but the combination of spectral, timing, and polarimetric data is probably the best tool in the box to determine once and for all whether a binary SMBH is indeed present in Mrk~231. A conclusive piece of evidence brought forth by our study would be the presence of a far-UV bump in the polarized flux of the AGN. This is because such a feature cannot be associated with starburst activity nor dust scattering - and only with thermal emission and reprocessing in the disk atmosphere of a SMBH companion. 


\section{Conclusions}
\label{Conclusions}

We explored the anomalous total flux SED and polarization properties of the FeLoBAL quasar Mrk~231 in the framework of the binary SMBH model proposed by \citet{Yan_2015}. Our model, which also accounts for the emission of the dusty torus and of the host starlight, simultaneously fits the observed total and polarized fluxes. It is also able to reproduce the two puzzling features of Mrk~231, namely, the $\sim$ 2275~\AA\, break and the extreme peak of linear continuum polarization at about 3000~\AA, together with smooth rotations of the polarization angle, without invoking anything else that a binary SMBH and basic AGN features. 

To do so, the core of Mrk~231 should host a binary system of SMBH with a mass of $M_{*,p} \approx$ 1.6 $\times$ 10$^{8} M_{\odot}$ and $M_{*,s} \approx$ 1.1 $\times$ 10$^{7} M_{\odot}$, separated by a semimajor axis of $a_{BBH} \sim$ 146 AU. The secondary SMBH (and its accretion disk) imposes the observed $\sim$ 3 \% polarization degree in the 1000-2500~\AA\, waveband, while the primary black hole, with its truncated accretion disk and powerful winds, is responsible for the polarization peak. Those winds produce a scattering-induced polarization degree of about 23\%, which is highly diluted by the host starlight emission. The dusty torus is merely polarized, while the host galaxy is not. The vectorial sum of each of the polarized components leads to the smooth polarization position angle rotations observed in the spectrum. 

This is the first time that a binary SMBH model is used to successfully reproduce the total flux, the polarization degree and the polarization angle all at once. This is a particularity that is essential because they are three independent measurements. As a by-product, our model predicts a second bump in polarized flux at far-UV wavelength, something that cannot be probed at present due to the lack of space-based far-UV spectropolarimeters. Nevertheless, this study provides a critical step toward the understanding of the complex structure of Mrk~231 and the role of binary SMBHs in the evolution of AGNs. It opens pathways for future investigations of the internal structure of quasars thanks to space-based polarimetry.


\begin{acknowledgements}
The authors would like to acknowledge the support of the CNES, the CNRS, the University of Strasbourg, the PNHE and the PNCG. FM and JB are thankful to Brent Groves for his insight onto starburst modeling.
\end{acknowledgements}

\bibliographystyle{aa}
\bibliography{biblio}

\begin{appendix} 

\onecolumn

\section{HST/FOC imaging of Mrk~231}
\label{app:FOCreduc}

Imaging polarimetry using the HST/FOC require multiple observations through three polarizer filters whose fast axis are oriented along 0$^\circ$, 60$^\circ$, and 120$^\circ$, respectively. To obtain a polarization map, multiple steps are required for proper reduction ; these are presented in \cite{Barnouin2023} and references therein. 

\begin{figure*}[h!]
    \includegraphics[width=\linewidth]{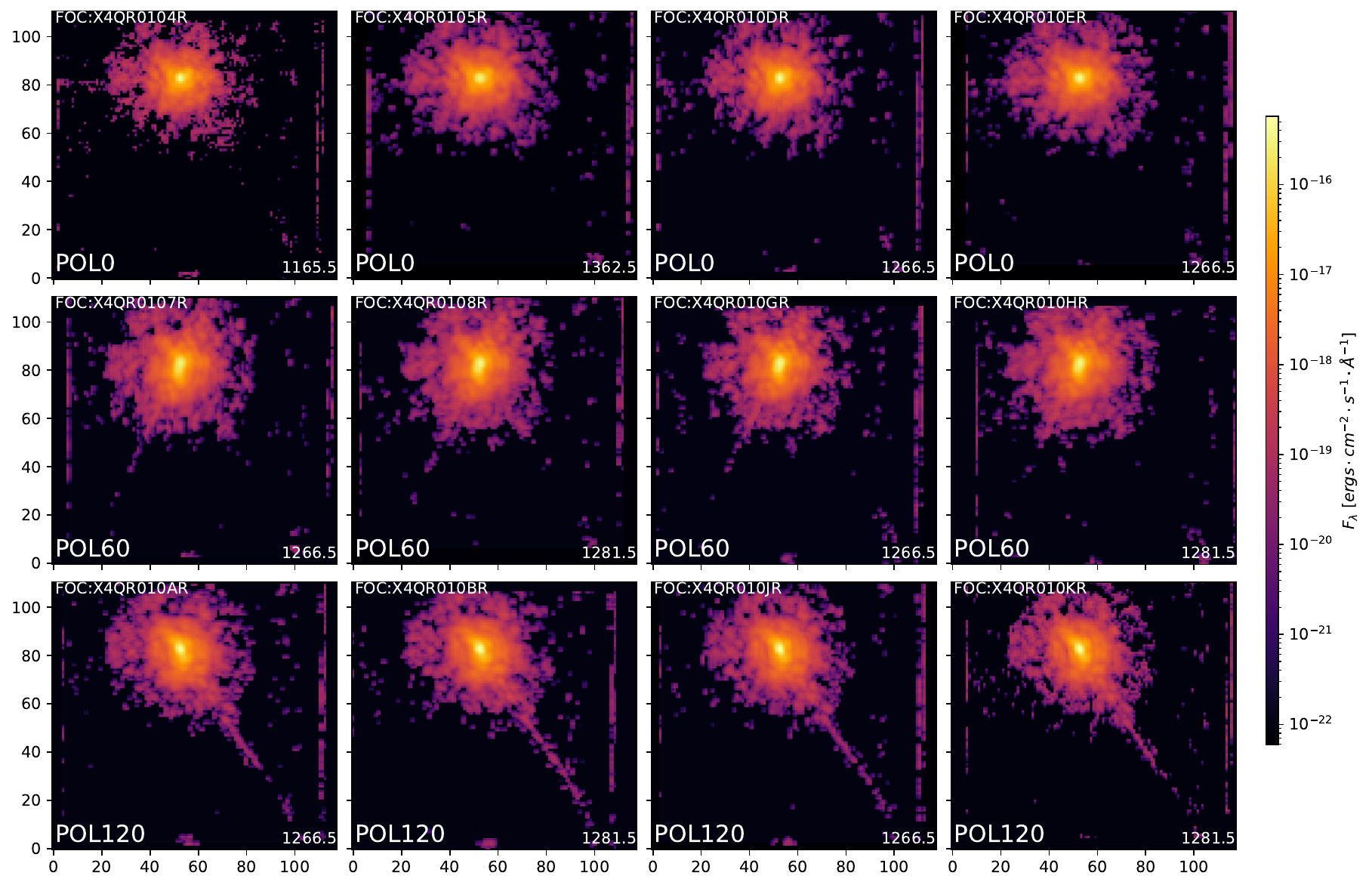}
    \caption{Full HST/FOC observation of Mrk~231 for ObsID 6444 as found in the archives. On each plate is displayed the dataset name (top left), the polarizer filter in use (bottom left) and the exposure time in second (bottom right). The X- and y-axes are in pixel offset.}
    \label{fig:FOC_obs}
\end{figure*}

As reported in \cite{Leighly2016}, each filter on the HST/FOC filter wheel has its own distinct PSF, that must be taken into account along the detector response when performing spatial analysis of the FOC imaging. We present in Fig. \ref{fig:FOC_obs} the flux and flatfield calibrated imaging for ObsID 6444 as taken from the Mast archives. Four observations have been taken through each polarizer filter to minimize detector noise while maximizing the signal to noise ratio. The detector readout format was set to 128x128 pixels to mitigate saturation. 

Upon study of the flux density at native resolution through each polarizer filter, we observe an extended feature that seem correlated with the polarizer filter axis, see Fig. \ref{fig:FOC_obs} (first row : 0$^\circ$, second row : 60$^\circ$ and third row : 120$^\circ$). This extended feature can result from saturation along the fast axis of the polarizer filter and cannot be corrected without the appropriate PSF taken at the time of the observations (Duccio Macchetto, private communication), so any spatial analysis of the polarization map for this observation is prevented.

\section{Results from Monte Carlo simulations of the BAL winds polarization}

Here, we show the outcomes from our extended study of the BAL winds polarization resulting from various wind optical thickness and composition (dust or electrons). 

\begin{figure}
    \centering
    \includegraphics[width=1\linewidth, trim={0cm 0cm 0cm 0cm}, clip]{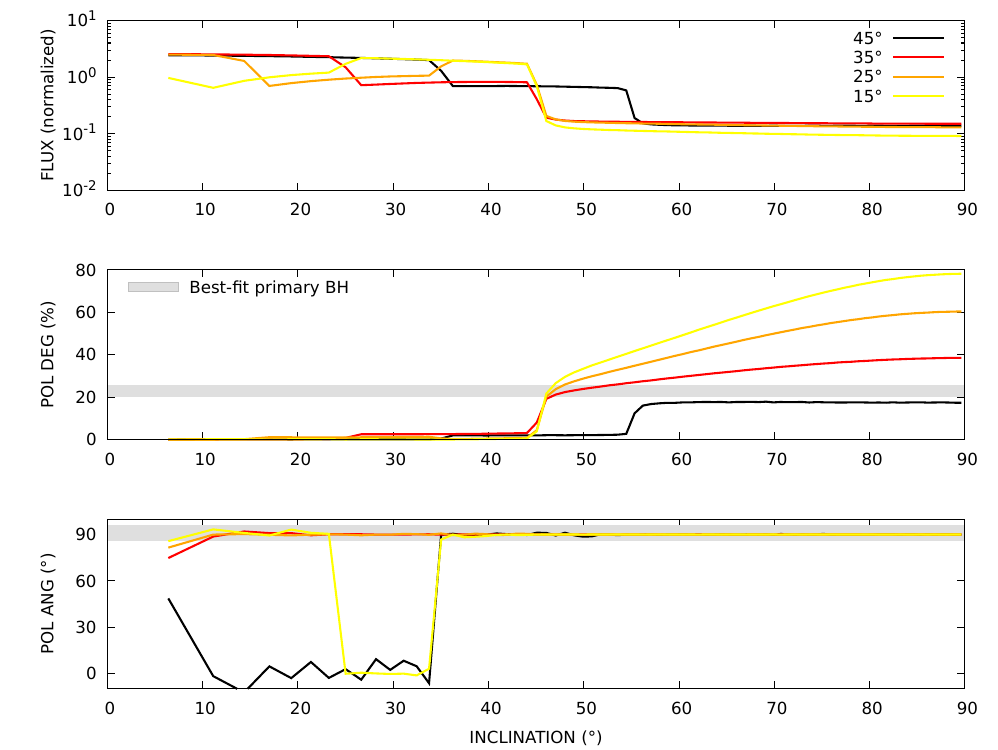}
    \caption{Total flux (normalized to the peak of emission, top), continuum linear polarization degree (middle) and polarization angle (bottom) as a function of the observer's viewing angle, defined with respect to the symmetry axis of the model, see Fig.~\ref{Fig:Model}. Four different inclinations of the wind are shown : 15$^\circ$, 25$^\circ$, 35$^\circ$ and 45$^\circ$. The wind is fully ionized and has an optical thickness of 1 along the outflowing direction. In gray is the polarization expected from the primary component of the SMBH binary found by our best-fit analysis.}
    \label{Fig:Sim_elec_1}
\end{figure}

\begin{figure}
    \centering
    \includegraphics[width=0.5\linewidth, trim={0cm 0cm 0cm 0cm}, clip]{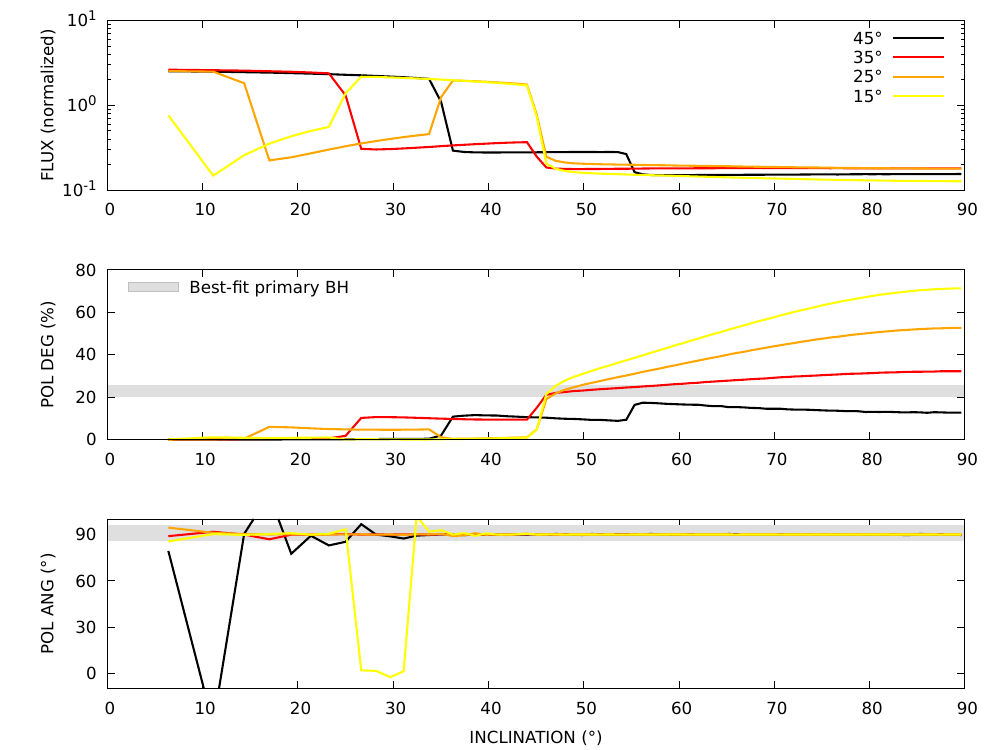}
    \caption{Same as Fig.~\ref{Fig:Sim_elec_1} but for an optical thickness of 2.5 along the outflowing direction.}
    \label{Fig:Sim_elec_2_5}
\end{figure}

Figs.~\ref{Fig:Sim_elec_1}, \ref{Fig:Sim_elec_2_5}, and \ref{Fig:Sim_elec_5} present the values obtained in the case of a wind filled with electrons, with different optical depths. One can see that, at a viewing angle that matches the outflow direction (where BAL are observed), the total flux decreases as photons are scattered away from the observer's viewing angle. The associated polarization degree rises, since the unpolarized emission source is obscured by the wind geometry and the polarization position angle is always perpendicular to the symmetry axis of the model (corresponding to the jet direction). However, the polarization behavior can be completely different when the observer's inclination is lower or higher than that of the outflow's direction. Once the observer's line-of-sight crosses the dusty torus, the polarization degree naturally rises up to several tens of percents and is always associated with a perpendicular polarization angle, as expected from AGN simulations \citep{Marin2012}. In the case of Thomson scattering, only winds with an optical thickness of 5 or more can reach the peak of polarization degree observed in Mrk~231, and only for large inclinations ($\sim$ 45$^\circ$) of the outflows.

Figs.~\ref{Fig:Sim_dust_1}, \ref{Fig:Sim_dust_2_5}, and \ref{Fig:Sim_dust_5} show the same results but for a wind that is dominated by Mie scattering. In this case, while similar behaviors are found for the evolution of the total flux as a function of the observer's line-of-sight, the associated polarization degree is much lower, of the order of a percent at maximum, rising to 6\% at edge-on inclinations. The polarization angle recorded when the observer's line-of-sight crosses the outflowing direction is orthogonal to the one measured in the literature. Both indicators (the polarization degree and polarization angle) rule out a dust-dominated wind model in Mrk~231 to explain the polarization peak observed in the optical. The reason why dusty winds produce lower and orthogonal polarization with respect to similar models filled with electrons is that electron scattering has an isotropic phase function in the optical, while Mie scattering disfavors forward scattering (towards the observer), where absorption will dominate. 

\begin{figure}
    \centering
    \includegraphics[width=0.5\linewidth, trim={0cm 0cm 0cm 0cm}, clip]{image/Electron_polarization_tau_5.pdf}
    \caption{Same as Fig.~\ref{Fig:Sim_elec_1} but for an optical thickness of 5 along the outflowing direction.}
    \label{Fig:Sim_elec_5}
\end{figure}

\begin{figure}
    \centering
    \includegraphics[width=0.5\linewidth, trim={0cm 0cm 0cm 0cm}, clip]{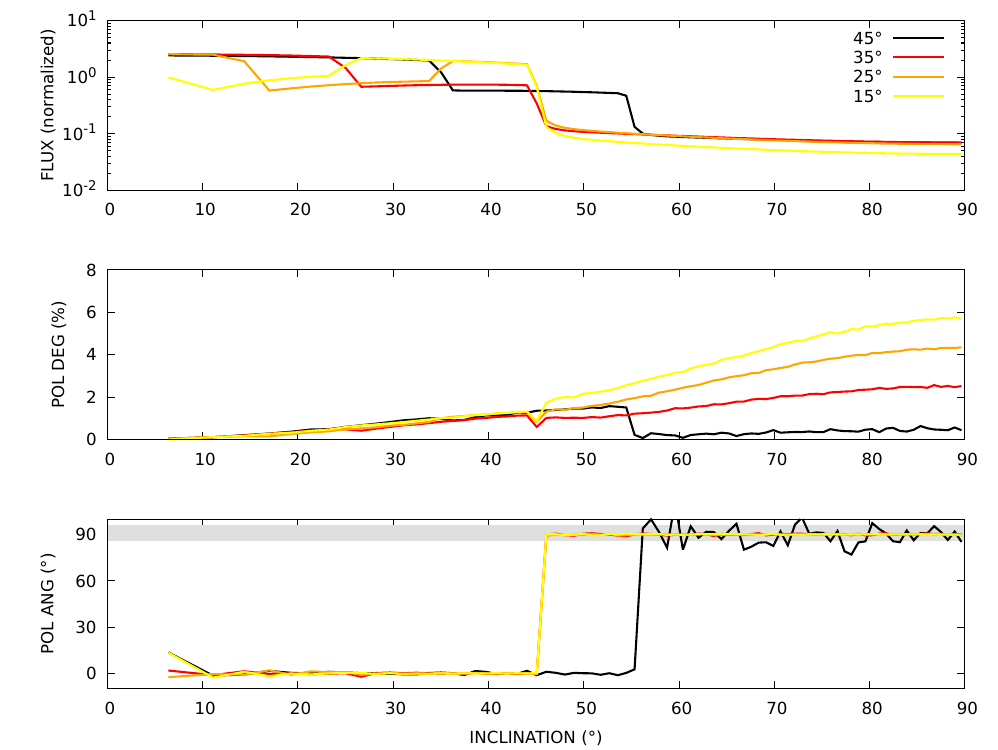}
    \caption{Same as Fig.~\ref{Fig:Sim_elec_1} but for a wind filled a Milky Way dust mixture that has an optical thickness of 1 along the outflowing direction.}
    \label{Fig:Sim_dust_1}
\end{figure}

\begin{figure}
    \centering
    \includegraphics[width=0.5\linewidth, trim={0cm 0cm 0cm 0cm}, clip]{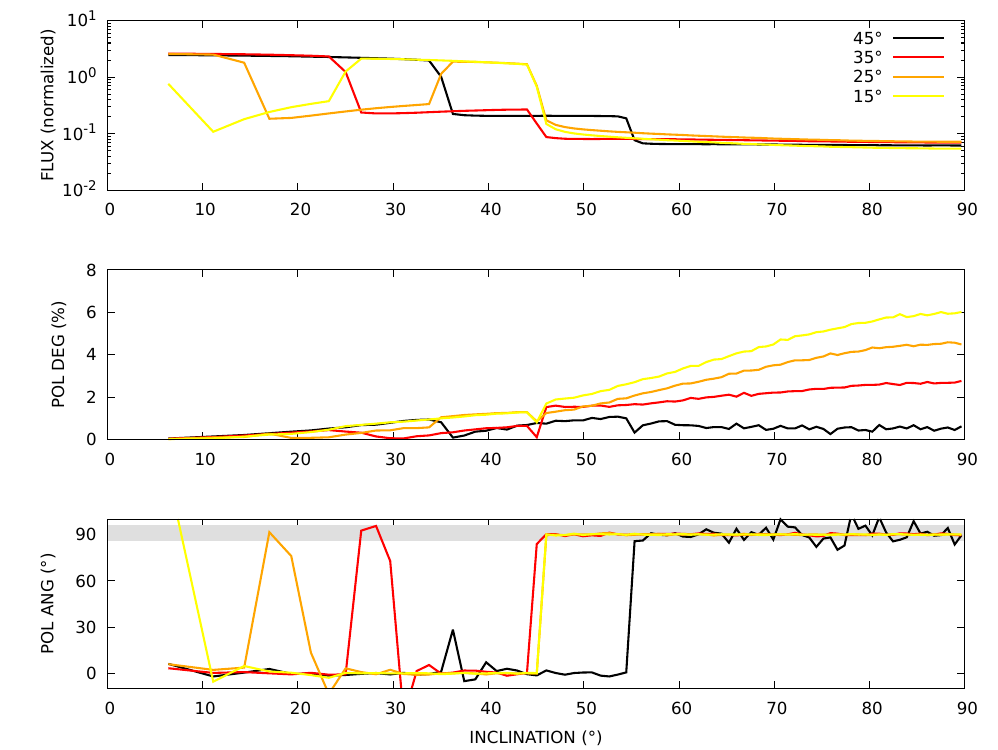}
    \caption{Same as Fig.~\ref{Fig:Sim_elec_1} but for a wind filled a Milky Way dust mixture that has an optical thickness of 2.5 along the outflowing direction.}
    \label{Fig:Sim_dust_2_5}
\end{figure}

\begin{figure}
    \centering
    \includegraphics[width=0.5\linewidth, trim={0cm 0cm 0cm 0cm}, clip]{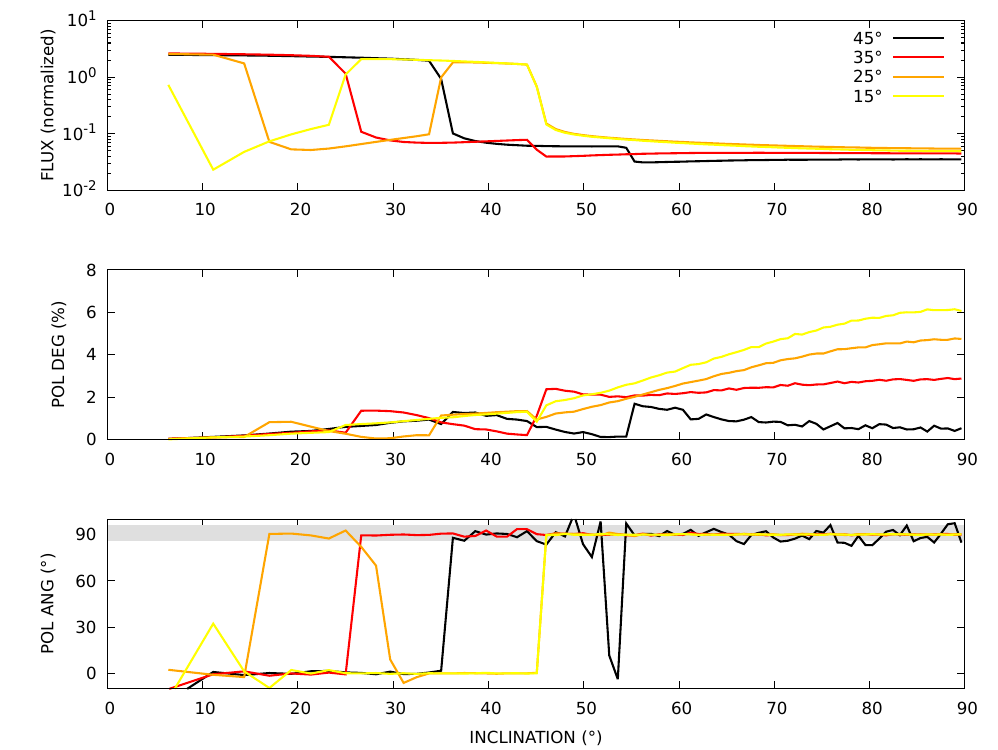}
    \caption{Same as Fig.~\ref{Fig:Sim_elec_1} but for a wind filled a Milky Way dust mixture that has an optical thickness of 5 along the outflowing direction.}
    \label{Fig:Sim_dust_5}
\end{figure}

\end{appendix}

\end{document}